\documentclass[11pt,a4paper]{article}
\pdfoutput=1
\usepackage{jcappub}
\usepackage{ifthen}
\usepackage{epsfig}
\usepackage{booktabs} 
\usepackage{natbib}

\usepackage{morefloats}

\newcommand{\beqra}{\begin{eqnarray}}
\newcommand{\eeqra}{\end{eqnarray}}
\newcommand{\beq}{\begin{equation}}
\newcommand{\eeq}{\end{equation}}
\title{Analysis of the theoretical bias in dark matter direct detection}
\author[a]{Riccardo Catena}
\affiliation[a]{Institut f\"ur Theoretische Physik, Friedrich-Hund-Platz 1, 37077 G\"ottingen, Germany}
\emailAdd{riccardo.catena@theorie.physik.uni-goettingen.de}

\abstract{Fitting the model ``A'' to dark matter direct detection data, when the model that underlies the data is ``B'', introduces a theoretical bias in the fit. We perform a quantitative study of the theoretical bias in dark matter direct detection, with a focus on assumptions regarding the dark matter interactions, and velocity distribution. We address this problem within the effective theory of isoscalar dark matter-nucleon interactions mediated by a heavy spin-1 or spin-0 particle. We analyze 24 benchmark points in the parameter space of the theory, using frequentist and Bayesian statistical methods. First, we simulate the data of future direct detection experiments assuming a momentum/velocity dependent dark matter-nucleon interaction, and an anisotropic dark matter velocity distribution. Then, we fit a constant scattering cross section, and an isotropic Maxwell-Boltzmann velocity distribution to the simulated data, thereby introducing a bias in the analysis. The best fit values of the dark matter particle mass differ from their benchmark values up to 2 standard deviations. The best fit values of the dark matter-nucleon coupling constant differ from their benchmark values up to several standard deviations. We conclude that common assumptions in dark matter direct detection are a source of potentially significant bias.}


\begin{document}
\maketitle

\section{Introduction}

Dark matter constitutes about five sixth of the total matter in the observable Universe~\cite{Ade:2013zuv}, and it forms large spheroidal halos hosting the majority of the astrophysical structures, including our Milky Way~\cite{Jungman:1995df,Bertone:2004pz}. 

The particles forming the Milky Way dark matter halo have so far escaped detection, and the dark matter mass, as well as its interactions, remain unknown. 
The detection of dark matter particles in the solar neighborhood through their scattering off the nuclei of a terrestrial detector would provide us with a direct measurement of the dark matter particle mass and of the strength of the dark matter-nucleon interaction~\cite{Strigari:2013iaa}. To accomplish this goal is the aim of the dark matter direct detection technique~\cite{Lewin:1995rx}.

The direct detection technique has experienced a great advancement in the past few years. Current experiments can probe spin-independent dark matter-nucleon scattering cross sections at the level of $10^{-45}$~cm$^2$~\cite{Aprile:2012nq,Ahmed:2009zw}, or below~\cite{Akerib:2013tjd}, for weakly interacting dark matter candidates of mass around 30~GeV, and cross sections of approximately $10^{-41}$~cm$^2$, for masses in the 5-10~GeV range~\cite{Agnese:2014aze,Aalseth:2014jpa}. The next generation of direct detection experiments will exploit target masses at the ton-scale, improving current exclusion limits on the dark matter-nucleon scattering cross-section by about two order of magnitudes~\cite{Akrami:2010dn,Pato:2010zk,Strege:2012kv,Cerdeno:2013gqa,Cerdeno:2014uga}.

Fitting the dark matter particle mass and coupling constants to direct detection data, assumptions about the dark matter-nucleon interaction and the local dark matter velocity distribution are necessary~\footnote{Alternatively, one can assume that the dark matter particle mass is known, relax the assumption on the dark matter velocity distribution, and compare different experiments in a halo independent approach~\cite{Drees:2008bv,Fox:2010bz,Fox:2010bu,McCabe:2011sr,Frandsen:2011gi,DelNobile:2013cva,Bozorgnia:2013hsa}.}~\cite{Green:2010gw}. Dark matter-nucleon scattering cross sections independent of the momentum transfer and of the relative velocity, and a Maxwell-Boltzmann velocity distribution are common assumptions in this field~\cite{2012arXiv1209.3339F}. The former assumption is motivated by the small velocity of the dark matter particles in the Milky Way, the latter by the simplicity of the Maxwell-Boltzmann distribution (i.e. a self-consistent distribution generated by the density profile of an isothermal sphere~\cite{GD}). Although both assumptions are well-motivated, other interaction types and velocity distributions are equally plausible, and could be considered in the data analysis~\cite{Catena:2013pka}. 

In the dark matter-nucleon scattering, momentum and velocity independent interaction operators arise when dark matter couples to the nuclear charge density operator, or to the nuclear spin current density operator~\cite{Fitzpatrick:2012ix}. The former operator generates the so-called spin-independent interaction, the latter the so-called spin-dependent interaction. Despite the small velocity of the dark matter particles in the Milky Way, these operators do not necessarily generate the leading contributions to the non relativistic limit of the dark matter-nucleus scattering cross section. There are interesting examples where momentum and velocity independent interaction operators are forbidden by symmetry arguments, or parametrically suppressed~\cite{Chang:2009yt,Fan:2010gt,McDermott:2011hx,Fitzpatrick:2012ib,Anand:2013yka,DelNobile:2013sia,Gresham:2014vja,DelNobile:2014eta,Panci:2014gga}. 

In studying the velocity distribution of dark matter particles in the solar neighborhood, various approaches have been investigated, including Markov Chain Monte Carlo analyses~\cite{Catena:2011kv,Fairbairn:2012zs,Bozorgnia:2013pua,Fornasa:2013iaa} and N-body simulations~\cite{Vogelsberger:2008qb,Kuhlen:2009vh,McCabe:2010zh}. As shown by the N-body simulations, the Maxwell-Boltzmann distribution only provides an approximate description of the particles forming the Milky Way dark matter halo. For instance, N-body simulations predict that the dark matter velocity distribution is anisotropic in a broad range of galactocentric distances, in contrast to the standard Maxwell-Boltzmann approximation~\cite{Ludlow:2011cs}. Interesting polynomials expansions~\cite{Kavanagh:2013wba}, and decompositions in streams~\cite{Feldstein:2014gza} to model general dark matter velocity distributions have been recently proposed.

In summary, standard assumptions regarding the dark matter interactions, and velocity distribution might oversimplify the interpretation of future direct detection experiments, or even be incorrect, thereby introducing a bias in the data analysis. In this paper, we perform a quantitative study of the theoretical bias in dark matter direct detection, with a focus on assumptions regarding the dark matter-nucleon interaction, and the dark matter velocity distribution. We address this problem within the effective theory of isoscalar dark matter-nucleon interactions mediated by a heavy spin-1 or spin-0 particle~\cite{Fan:2010gt,Fitzpatrick:2012ix}. 

First, we simulate future dark matter direct detection data assuming momentum and velocity dependent dark matter-nucleon interactions, and an anisotropic dark matter velocity distribution. Then, we analyze the simulated data assuming either a constant dark-matter nucleon scattering cross section, or an isotropic Maxwell-Boltzmann velocity distribution, hence introducing a bias in the analysis. Comparing the best fit points with the original benchmark points, we estimate the bias in dark matter direct detection potentially induced by incorrect theoretical assumptions. 

We present the first study of the bias in dark matter direct detection based on the general effective theory of isoscalar dark matter-nucleon interactions. We cover both particle physics and astrophysical aspects of the problem. Previous studies restrict the analysis to the familiar spin-independent and spin-dependent interactions~\cite{Cerdeno:2012ix}, or to a bias of astrophysical nature~\cite{Pato:2012fw,Peter:2013aha}. 

The paper is organized as follows. In Sec.~\ref{sec:theory}, we briefly review the theoretical framework and the statistical methods adopted in the analyses. Sec.~\ref{sec:analysis} is devoted to a description of the simulated data. In Sec.~\ref{sec:results} we present our results. We conclude in Sec.~\ref{sec:conclusions}. Useful dark matter response functions are listed in the appendix. 

\section{Theoretical background}
\label{sec:theory}
\subsection{Dark matter-nucleon effective theory}
The most general Lagrangian describing the dark matter-nucleon interactions arising from the exchange of a heavy spin-0 or spin-1 particle is given by the linear combination 
\begin{equation}
\mathcal{L}_{\rm int} = \sum_{N={\rm n}, {\rm p}}\sum_{i} c_i^{N} \mathcal{O}_i \chi^+ \chi^- N^+ N^- \,,
\label{eq:L}
\end{equation}
where $\chi^+$ ($\chi^-$) and $N^+$ ($N^-$) are the positive (negative) frequency parts of the non-relativistic dark matter and nucleon fields, respectively. The operators $\mathcal{O}_i$, with $i=1,3,\dots,11$, are restricted by Galilean invariance, energy and momentum conservation and hermiticity. They are listed in Tab.~\ref{tab:operators}. The interaction operators $\mathcal{O}_1$ and $\mathcal{O}_4$ generate the familiar spin-independent and spin-dependent interactions. In Eq.~(\ref{eq:L}), $c^{\rm p}_i$ and $c^{\rm n}_i$ denote the coupling constants for protons and neutrons, respectively. Their linear combinations  
\begin{equation}
c^{0}_i = c^{\rm p}_i+c^{\rm n}_i\,; \qquad \qquad c^{1}_i = c^{\rm p}_i-c^{\rm n}_i\,,  
\label{eq:couplings}
\end{equation}
define the isoscalar ($c^{0}_i$) and isovector ($c^{1}_i$) coupling constants. In this paper we restrict our investigations to isoscalar interactions, i.e., we set $c^{1}_i=0$. We collectively denote the isoscalar coupling constants by $\mathbf{c}$. Following Ref.~\cite{Anand:2013yka}, we define the coupling constants $c^0_i$ with dimension (mass)$^{-2}$. An interesting example in which isovector couplings are not negligible is studied in Ref.~\cite{Arina:2014yna}. In this model, dark matter couples to protons with a larger strength as compared to neutrons, which implies a number of important phenomenological consequences. For instance, dark matter is expected to scatter off target nuclei with unpaired protons more likely than target nuclei with unpaired neutrons.
\begin{table}[t]
    \centering
    \begin{tabular}{ll}
    \toprule
        $\mathcal{O}_1 = 1_{\chi} 1_{N}$ \hspace{10em} &         $\mathcal{O}_7 = \vec{S}_{N}\cdot \vec{v}^{\perp}_{\chi N}$ \\
        $\mathcal{O}_3 = -i\vec{S}_N\cdot\left(\frac{\vec{q}}{m_N}\times\vec{v}^{\perp}_{\chi N}\right)$ &         $\mathcal{O}_8 = \vec{S}_{\chi}\cdot \vec{v}^{\perp}_{\chi N}$ \\
        $\mathcal{O}_4 = \vec{S}_{\chi}\cdot \vec{S}_{N}$ &         $\mathcal{O}_9 = -i\vec{S}_\chi\cdot\left(\vec{S}_N\times\frac{\vec{q}}{m_N}\right)$ \\                                                                             
        $\mathcal{O}_5 = -i\vec{S}_\chi\cdot\left(\frac{\vec{q}}{m_N}\times\vec{v}^{\perp}_{\chi N}\right)$ &         $\mathcal{O}_{10} = -i\vec{S}_N\cdot\frac{\vec{q}}{m_N}$ \\                                                                                                        
        $\mathcal{O}_6 = \left(\vec{S}_\chi\cdot\frac{\vec{q}}{m_N}\right) \left(\vec{S}_N\cdot\frac{\vec{q}}{m_N}\right)$ &        $\mathcal{O}_{11} = -i\vec{S}_\chi\cdot\frac{\vec{q}}{m_N}$ \\                                                                                                      
    \bottomrule
    \end{tabular}
    \caption{Non-relativistic operators appearing in Eq.~(\ref{eq:L}). The operators $\mathcal{O}_i$ are the same as in Ref.~\cite{Anand:2013yka}. We adopt the following notation: $1_\chi 1_N$ is the identity, $\vec{q}$ is the momentum transfer, $\vec{v}^{\perp}_{\chi N}$ is the $\chi$-nucleon transverse relative velocity operator, $\vec{S}_\chi 1_N$ and $1_\chi \vec{S}_{N}$ are the dark matter and nucleon spin operators, respectively. $\chi$ denotes the dark matter particle.}
    \label{tab:operators}
\end{table}

From the Lagrangian in Eq.~(\ref{eq:L}), we calculate the expected number of dark matter scattering events in a target material. The differential rate of scattering events per unit time and per unit detector mass is given by
\begin{equation}
\frac{{\rm d}\mathcal{R}}{{\rm d}E_{R}} =  \sum_{T}\frac{{\rm d}\mathcal{R}_{T}}{{\rm d}E_{R}} \equiv  \sum_{T} \xi_T \frac{\rho_{\chi}}{2\pi m_\chi}  \left\langle  \frac{1}{v} P_{\rm tot}(v^2,q^2)  \right\rangle
\label{rate_theory}
\end{equation}
where $\rho_\chi$ is the local dark matter density, $m_\chi$ is the dark matter particle mass, and $\xi_T$ is the mass fraction of the nucleus $T$ in the target material. In Eq.~(\ref{rate_theory}), $P_{\rm tot}$ is the square modulus of the transition amplitude in the non-relativistic limit, i.e. $|\mathcal{M}_{NR}|^2$, averaged over initial spins and summed over final spins. $P_{\rm tot}$ can be written as a combination of nuclear and dark matter response functions:
\allowdisplaybreaks
\begin{eqnarray}
P_{\rm tot}({v}^2,{q}^2)&\equiv&{1 \over 2j_\chi + 1} {1 \over 2j_N + 1} \sum_{\rm spins} |\mathcal{M}_{NR}|^2 \nonumber \\  &=& {4 \pi \over 2j_N + 1} 
\sum_{ \tau=0,1} \sum_{\tau^\prime = 0,1} \Bigg\{ \Bigg[ R_{M}^{\tau \tau^\prime}({v}^{\perp 2}_{\chi T}, {{q}^{2} \over m_N^2})~W_{M}^{\tau \tau^\prime}(y)   \nonumber\\
&+& R_{\Sigma^{\prime \prime}}^{\tau \tau^\prime}({v}^{\perp 2}_{\chi T}, {{q}^{2} \over m_N^2})   ~W_{\Sigma^{\prime \prime}}^{\tau \tau^\prime}(y) 
+   R_{\Sigma^\prime}^{\tau \tau^\prime}({v}^{\perp 2}_{\chi T}, {{q}^{2} \over m_N^2}) ~ W_{\Sigma^\prime}^{\tau \tau^\prime}(y) \Bigg]  \nonumber\\  
&+& {{q}^{2} \over m_N^2} ~\Bigg[R_{\Phi^{\prime \prime}}^{\tau \tau^\prime}({v}^{\perp 2}_{\chi T}, {{q}^{2} \over m_N^2}) ~ W_{\Phi^{\prime \prime}}^{\tau \tau^\prime}(y)  +  R_{ \Phi^{\prime \prime}M}^{\tau \tau^\prime}({v}^{\perp 2}_{\chi T}, {{q}^{2} \over m_N^2})  ~W_{ \Phi^{\prime \prime}M}^{\tau \tau^\prime}(y) \nonumber\\
&+&   R_{\tilde{\Phi}^\prime}^{\tau \tau^\prime}({v}^{\perp 2}_{\chi T}, {{q}^{2} \over m_N^2}) ~W_{\tilde{\Phi}^\prime}^{\tau \tau^\prime}(y) 
+   R_{\Delta}^{\tau \tau^\prime}({v}^{\perp 2}_{\chi T}, {{q}^{2} \over m_N^2}) ~ W_{\Delta}^{\tau \tau^\prime}(y) \nonumber\\
 &+&  R_{\Delta \Sigma^\prime}^{\tau \tau^\prime}({v}^{\perp 2}_{\chi T}, {{q}^{2} \over m_N^2})  ~W_{\Delta \Sigma^\prime}^{\tau \tau^\prime}(y)   \Bigg]  \Bigg\}  \,,
\label{Ptot}
\end{eqnarray}
where
\begin{align}
v^{\perp 2}_{\chi T} = v^2 - \frac{q^2}{4\mu_T^2} \,.
\end{align}
In Eq.~(\ref{Ptot}), $v$ and $\mu_T$ are the dark matter-nucleus relative velocity and reduced mass, respectively, whereas $y=(q b/2)^2$, $q$ is the momentum transfer, and $b$ is the oscillator parameter in the independent-particle harmonic oscillator model~\cite{Anand:2013yka}. $j_N$ and $j_\chi$ are the nucleus spin and the dark matter particle spin, respectively. For definiteness, we assume  that the dark matter particle has spin $j_\chi=1/2$. 

The nuclear response functions  $W_{M}^{\tau \tau^\prime}$, $W_{\Sigma^{\prime \prime}}^{\tau \tau^\prime}$, $W_{\Sigma^\prime}^{\tau \tau^\prime}$, $W_{\Phi^{\prime \prime}}^{\tau \tau^\prime}$, $W_{\Phi^{\prime\prime}M}^{\tau \tau^\prime}$, $W_{\tilde{\Phi}^\prime}^{\tau \tau^\prime}$, $W_{\Delta}^{\tau \tau^\prime}$ and $W_{\Delta \Sigma^\prime}^{\tau \tau^\prime}$ are defied in Eq.~(41) of Ref.~\cite{Anand:2013yka}. We evaluate them using our {\sffamily FORTRAN} version of the {\sffamily Mathematica} package described in Ref.~\cite{Anand:2013yka}. The dark matter response functions $R_{M}^{\tau \tau^\prime}$, $R_{\Sigma^{\prime \prime}}^{\tau \tau^\prime}$, $R_{\Sigma^\prime}^{\tau \tau^\prime}$, $R_{\Phi^{\prime \prime}}^{\tau \tau^\prime}$, $R_{\Phi^{\prime\prime}M}^{\tau \tau^\prime}$, $R_{\tilde{\Phi}^\prime}^{\tau \tau^\prime}$, $R_{\Delta}^{\tau \tau^\prime}$ and $R_{\Delta \Sigma^\prime}^{\tau \tau^\prime}$ are given in the Appendix. 

In Eq.~(\ref{rate_theory}), the angle brackets denote the average
\begin{equation}
\left\langle \frac{1}{v} P_{\rm tot}(v^2,q^2)  \right\rangle =  \int\limits_{v>v_{\rm min}(q)} \,  \frac{f(\vec{v} + \vec{v}_e(t))}{v} \, P_{\rm tot}(v^2,q^2) \, d^3v  , 
\label{eq:uDF}
\end{equation} 
where $v_{\rm min}(q)=q/2\mu_T$ is the minimum velocity that a dark matter particle must have in order to transfer a momentum $q$ to the target nucleus, and $f$ is the local dark matter velocity distribution in the galactic rest frame boosted to the detector frame. In Eq.~(\ref{eq:uDF}), $\vec{v}_e(t)$ is the time-dependent Earth velocity in the galactic rest frame. 

In the simulations of Sec.~\ref{sec:analysis}, we assume the self-consistent anisotropic velocity distribution proposed in Ref.~\cite{Bozorgnia:2013pua}. This velocity distribution has been obtained from a generalized Eddington's inversion formula applied to the galactic model of Refs~\cite{Catena:2009mf,Catena:2011kv}. We set the relevant astrophysical parameters at their mean values, i.e. blue line in the left panel of Fig.~6 in Ref.~\cite{Bozorgnia:2013pua}. In Sec.~\ref{sec:results2}, we also use a Maxwell-Boltzmann distribution $f(\vec{v} + \vec{v}_e(t))\propto \exp(-|\vec{v}+\vec{v}_{e}(t)|^2/v_0^2)$ truncated at the local escape velocity $v_{\rm esc}=544$~km~s$^{-1}$, with $v_{0}=220$~km~s$^{-1}$ and $\rho_\chi=0.3$~GeV~cm$^{-3}$ (i.e. the standard dark matter halo).

\subsection{Statistical framework}
\label{sec:statistics}
We now introduce the statistical methods that we will use in Sec.~\ref{sec:results} in order to analyze the synthetic data introduced in Sec.~\ref{sec:analysis}. We adopt frequentist and Bayesian statistical methods based on the likelihood function, $\mathcal{L}(\mathbf{d} | \mathbf{\Theta})$. Here $\mathbf{d}$ denotes an array of simulated data. $\mathbf{\Theta}~\equiv~(\theta_1,\theta_2)$ is a 2-dimensional array composed of $m_\chi$ and of one of the coupling constants $c_i^0$. The index $i$ depends on the simulated data that we analyze (see Secs.~\ref{sec:analysis} and \ref{sec:results} for concrete applications). The likelihood function that we use in the analyses is defined in Sec.~\ref{sec:likelihood}.

In the frequentist approach, we use the likelihood function to construct an effective chi-square defined as $\Delta \chi^2_{\rm eff}\equiv-2 \ln \mathcal{L}/ \mathcal{L}_{\rm max}$, where $\mathcal{L}_{\rm max}$ is the absolute maximum of the likelihood function. From $\Delta \chi^2_{\rm eff}$, we derive approximate 2D frequentist confidence intervals, which admit a rigorous statistical interpretation when the regularity conditions underlying Wilk's theorem apply. We maximize the likelihood function to obtain the best fit points of the model parameters.

In the Bayesian approach, we calculate the posterior probability density function (PDF) of the model parameters $\mathcal{P}(\mathbf{\Theta} | \mathbf{d})$, applying Bayes' theorem to the likelihood function $\mathcal{L}(\mathbf{d} | \mathbf{\Theta})$, 
\begin{equation}
\mathcal{P}(\mathbf{\Theta} | \mathbf{d}) = \mathcal{L}(\mathbf{d} | \mathbf{\Theta}) \pi(\mathbf{\Theta})\mathcal{E}(\mathbf{d})^{-1} \,.
\label{eq:bayes}
\end{equation} 
In Eq.~(\ref{eq:bayes}), $\pi(\mathbf{\Theta})$ is the prior PDF. Any prior information on the model parameters must be included in this term. In absence of any empirical guideline, we assume log-priors for $m_\chi$ and $\mathbf{c}$. Tab.~\ref{tab:priors} shows the model parameters and the support of the associated prior PDFs. $\mathcal{E}(\mathbf{d})$ is the evidence, and in the present analysis it plays the role of a normalization constant. From the posterior PDF, we derive the posterior means of the model parameters. The posterior mean of the parameter $\theta_1$, for instance, is defined as follows 
\begin{equation}
\langle \theta_1 \rangle = \int d \theta_{1}d \theta_{2}\, \theta_1  \mathcal{P}(\mathbf{\Theta} | \mathbf{d})  \,.
\label{eq:marg}
\end{equation}
Bayesian analog of the 2D frequentist confidence intervals are the 2D credible regions. 95\% 2D credible regions, for instance, contain the 95\% of the total posterior probability in the $\theta_1$-$\theta_2$ plane. By construction, $\mathcal{P}(\theta^{\rm in}_1, \theta^{\rm in}_2|\mathbf{d}) \ge \mathcal{P}(\theta^{\rm out}_1, \theta^{\rm out}_2|\mathbf{d})$, for ($\theta^{\rm in}_1$, $\theta^{\rm in}_2$) inside the credible region, and ($\theta^{\rm out}_1$, $\theta^{\rm out}_2$) outside the credible region. 

We define as ``bias parameter'' $\beta$ of the variable $X$, the absolute value of the difference between the best fit value of $X$, i.e. $X_{\rm bf}$, and the value of X at the benchmark point, i.e. $X_{\rm P}$, divided by the error $\Delta$:    
\begin{equation}
\beta = \frac{|X_{\rm bf}-X_{\rm P}|}{\Delta} \,.
\label{eq:}
\end{equation}
$\Delta$ is defined as half of the width of the 95\% confidence interval of the variable $X$. 

In Sec.~\ref{sec:analysis} we provide a detailed description of the likelihood function used in the analyses. We take advantage of the {\sffamily Multinest} program~\cite{Feroz:2008xx,Feroz:2007kg,Feroz:2013hea} with parameters set at $n_{\rm live}=20000$ and tol=$10^{-4}$ in order to sample the likelihood function. We use our own routines to calculate direct detection scattering rates and evaluate the likelihood function. Finally, we use the programs {\sffamily Getplots}~\cite{Austri:2006pe} and {\sffamily Matlab} to produce the figures.

\begin{table}
    \centering
    \begin{tabular}{lclc}
    \toprule
    Parameter         & Type & Prior range &  Prior type \\
    \midrule                                          
    $\log_{10} (m_v^2 c_1^0)$ & model parameter & $[-5,1]$ & log-prior  \\
    $\log_{10} (m_v^2c_3^0)$ & model parameter & $[-4,4]$ & log-prior  \\
    $\log_{10} (m_v^2c_4^0)$ & model parameter & $[-2,3]$ & log-prior  \\
    $\log_{10} (m_v^2c_5^0)$ & model parameter & $[-4,4]$ & log-prior  \\
    $\log_{10} (m_v^2c_6^0)$ & model parameter & $[-4,4]$ & log-prior  \\
    $\log_{10} (m_v^2c_7^0)$ & model parameter & $[-4,4]$ & log-prior  \\
    $\log_{10} (m_v^2c_8^0)$ & model parameter & $[-4,4]$ & log-prior  \\
    $\log_{10} (m_v^2c_9^0)$ & model parameter & $[-4,4]$ & log-prior  \\
    $\log_{10} (m_v^2c_{10}^0)$ & model parameter & $[-4,4]$ & log-prior  \\
    $\log_{10} (m_v^2c_{11}^0)$ & model parameter & $[-4,4]$ & log-prior  \\
    $\log_{10} (m_{\chi}/{\rm GeV})$ & model parameter & $[0.1,3 (4)]$  & log-prior  \\
    \bottomrule
    \end{tabular}
    \caption{List of model parameters. For each parameter, this table shows the support of the associated prior PDF, i.e. the prior range. As in~\cite{Anand:2013yka}, we express the coupling constants in units of $m_v^{-2}~=~(246.2~{\rm GeV})^{-2}$. For the benchmark points $P_{9}$ -- $P_{24}$ we extend the prior range of $\log_{10}(m_\chi/{\rm GeV})$ to 4.}
    \label{tab:priors}
\end{table}

\section{Analysis}
\label{sec:analysis}
This section is devoted to the simulation of our synthetic direct detection data. First, we describe the ton-scale detectors adopted in the simulations (Sec.~\ref{sec:detectors}), then we explain how to simulate synthetic data from a given detector and a fiducial point in the parameter space of the theory defined in Sec.~\ref{sec:theory} (Sec.~\ref{sec:simulations}). We analyze these data in Sec.~\ref{sec:results}, using the frequentist and Bayesian statistical methods described in Sec.~\ref{sec:statistics}.  

\subsection{Ton-scale detectors}
\label{sec:detectors}
In the analyses we assume two types of ton-scale detectors. A Germanium detector and a Xenon detector. For the ton-scale Germanium detector, we calculate the differential rate of dark matter scattering events per unit time and per unit detector mass as follows 
\begin{equation}
\frac{{\rm d}\mathcal{R}^{(1)}}{{\rm d}E_{\mathcal{O}}}=  \mathcal{E}\int_{0}^{\infty} dE_{R} \frac{1}{\sqrt{2\pi\sigma^2}} \exp\left[-\frac{(E_{R}-E_{\mathcal{O}})^2}{2\sigma^2}\right]  \frac{{\rm d}\mathcal{R}}{{\rm d}E_{R}} \,,
\label{eq:rate-Ge}
\end{equation}   
where~\cite{Akrami:2010dn} 
\begin{equation}
\sigma = \sqrt{\left(0.293\right)^2 + \left(0.056\right)^2 \left(E_R/ {\rm keV}\right)} \,
\end{equation}
is the energy dependent energy resolution of the detector, and $\mathcal{E}=0.3$ is the constant experimental efficiency. In Eq.~(\ref{eq:rate-Ge}), $E_{\mathcal{O}}$ is the observed energy, $E_{R}$ is the true nuclear recoil energy, and ${\rm d}\mathcal{R}/{\rm d}E_{R}$ is the the time average of the differential rate defined in Eq.~(\ref{rate_theory}). We calculate the total number of scattering events in the signal region ($E_{\rm inf}$, $E_{\rm sup}$) as
\begin{align}
\mu_S^{(1)}(m_\chi,\mathbf{c}) = MT\int_{E_{\rm inf}}^{E_{\rm sup}} \, \frac{{\rm d}\mathcal{R}}{{\rm d}E_{\mathcal{O}}} \, {\rm d}E_{\mathcal{O}} ,
\label{eq:muS}
\end{align}
where $MT=1000\times365$ kg-day is the raw exposure of the detector, $E_{\rm inf}=10$~keV and $E_{\rm inf}=100$~keV. Lower energy thresholds are possible, though they will be likely explored in a second stage of the data analysis~\cite{Cushman:2013zza}. We leave an analysis of this case for future work. The choice of a 10 keV energy threshold justifies our assumption of constant experimental efficiency (see Fig.~2 in Ref.~\cite{Agnese:2014aze}, and Ref.~\cite{Akrami:2010dn}).

For the ton-scale Xenon detector, we calculate the differential spectrum of dark matter induced photoelectrons (PE), $S_1$, as follows
\begin{equation}
\frac{{\rm d}\mathcal{R}^{(2)}}{{\rm d}S_1} = \mathcal{E}(S_1)\sum_{n=1}^{+\infty} {\rm Gauss}(S_1 | n,\sqrt{n}\sigma_{\rm PMT})  \int_{0}^{\infty}{\rm d}E_{R} \,{\rm Poiss}(n|\nu(E_{R})) \frac{{\rm d}\mathcal{R}}{{\rm d}E_{R}} \,.
\label{eq:rate_Xe}
\end{equation}
In Eq.~(\ref{eq:rate_Xe}), the Gaussian of standard deviation $\sqrt{n}\sigma_{\rm PMT}$, with $\sigma_{\rm PMT}=0.37$, and mean $n$ gives the probability of observing $S_1$ PE, when $n$ PE have been actually produced. The Poisson distribution of mean $\nu(E_R)$ gives the probability of producing $n$ PE from a recoil energy $E_R$. For the Xenon detector we assume the same $\nu(E_R)$, energy resolution, nuclear recoil acceptance, and efficiency of LUX~\cite{Akerib:2013tjd}. The function $\nu(E_R)$ can be extracted from Fig.~4 of Ref.~\cite{Akerib:2013tjd}. We set it to zero below 3 keV. In addition, we consider an exposure of $MT~=~1000~\times~365$ kg-day. In Eq.~(\ref{eq:rate_Xe}) ${\rm d}\mathcal{R}/{\rm d}E_{R}$ denotes the the time average of the differential rate defined in Eq.~(\ref{rate_theory}). We calculate the total number of scattering events in the signal region ($S_{1}^{\rm inf}$, $S_{1}^{ \rm sup}$) as follows
\begin{align}
\mu^{(2)}_S(m_\chi,\mathbf{c}) = MT\int_{S_{1}^{\rm inf}}^{S_{1}^{\rm sup}} \, \frac{{\rm d}\mathcal{R}}{{\rm d}S_{1}} \,{\rm d}S_1\,,
\label{eq:muS}
\end{align}
where $S_{1}^{\rm inf}=2$~PE and $S_{1}^{\rm sup}=30$~PE.

For both detector types we assume the spectrum of irreducible background events, 
\begin{equation}
\frac{{\rm d}\mathcal{R}^{(j)}_{\rm B}}{{\rm d}\hat{E}_j} = \frac{\eta}{b_j-a_j} +  \frac{\eta}{\epsilon_j\left[ \exp(-a_j/\epsilon_j)-\exp(-b_j/\epsilon_j)\right]} e^{-\hat{E}_j/\epsilon_j} \,,
\label{eq:background}
\end{equation}
which includes a flat component and an exponentially decreasing component~\cite{Akrami:2010dn}. In Eq.~(\ref{eq:background}), the index $j$ identifies the detector type. $j=1$ refers to the Germanium detector, whereas $j=2$ to the Xenon detector. Accordingly, $\hat{E}_1=E_\mathcal{O}$, $\hat{E}_2=S_1$, $a_1=10$ keV, $b_1=100$ keV, $\epsilon_1=10$ keV, $a_2=2$ PE, $b_2=30$ PE and, finally, $\epsilon_2=2$ PE. For both detectors, we assume one background event in the signal region, which implies $\eta=0.5$ in Eq.~(\ref{eq:background}). Accordingly, for the detector of type $j$, we define the total number of expected events in the signal region as
 \begin{equation}
\mu_{\rm tot}^{(j)}(m_{\chi},{\mathbf{c}}) = \mu^{(j)}_{S}(m_{\chi},{\mathbf{c}}) + 1\,. 
\end{equation}

In the calculations, we include the most abundant isotopes for each target material, namely: $^{70}$Ge, $^{72}$Ge, $^{73}$Ge, $^{74}$Ge and $^{76}$Ge, for Germanium and $^{128}$Xe, $^{129}$Xe, $^{130}$Xe, $^{131}$Xe, $^{132}$Xe, $^{134}$Xe, and $^{136}$Xe for Xenon. We calculate the nuclear response functions of the different isotopes using our {\sffamily FORTRAN} version of the {\sffamily Mathematica} package introduced in Ref.~\cite{Anand:2013yka}. 

\begin{table}
    \centering
    \begin{tabular}{cclcccc}
    \toprule
    Benchmark point   & $m_{\chi}$ (GeV) & $c_i^0\neq 0$  &  $N_1^{\rm th}$ & $N_1$ & $N_2^{\rm th}$& $N_2$  \\
    \midrule                                          
    $P_{1}$ & 10 & $m_v^2 c_3^0 = 20 $ & 1.7 & 3 &  86.3  & 81 \\
    $P_{2}$ & 10 & $m_v^2 c_5^0 = 120 $ &1.2 & 2 & 39.8 & 56  \\
    $P_{3}$ & 10 & $m_v^2 c_6^0 = 3000 $ & 2 & 3 & 75.9 & 71  \\
    $P_{4}$ & 10 & $m_v^2 c_7^0 = 2600 $ & 1.2 & 2 & 78 & 101 \\
    $P_{5}$ & 10 & $m_v^2 c_8^0 = 6$ & 1.4 & 2 & 87.3 & 82 \\
    $P_{6}$ & 10 & $m_v^2 c_9^0 = 80 $ & 1.8 & 3 & 73.8 & 96 \\
    $P_{7}$ & 10 & $m_v^2 c_{10}^0 = 50 $ & 1.7 & 3 & 69.9& 65  \\
    $P_{8}$ & 10 & $m_v^2 c_{11}^0 = 0.1$ & 1.5 & 2 & 48.9 & 67 \\
     \midrule  
    $P_{9}$ & 50 & $m_v^2 c_3^0 = 0.8 $ & 17.6 & 28 & 58.5 & 54 \\
    $P_{10}$ & 50& $m_v^2 c_5^0 = 12 $ & 34.4 & 31 & 75.5 & 98 \\
    $P_{11}$ & 50& $m_v^2 c_6^0 = 100 $ & 16.6 & 27 &  39.2 & 36 \\
    $P_{12}$ & 50 & $m_v^2 c_7^0= 300 $ & 17.9 & 29 & 52.8 & 49 \\
    $P_{13}$ & 50 & $m_v^2 c_8^0 = 0.5$ & 15.6 & 26 & 41.3 & 58 \\
    $P_{14}$ & 50 & $m_v^2 c_9^0 = 8$ &  24.8 & 38 & 59.1 & 79 \\
    $P_{15}$ & 50 & $m_v^ 2c_{10}^0 = 4$ & 20.1  & 32 & 60.2 & 56 \\
    $P_{16}$ & 50 & $m_v^2 c_{11}^0 = 0.01$ & 22.2 & 34 & 54.9 & 74 \\
     \midrule  
    $P_{17}$ & 250 & $m_v^2 c_3^0$ = 0.8 & 19.8 & 31 & 22.5 & 35 \\
    $P_{18}$ & 250& $m_v^2 c_5^0$ = 12 & 29.6 & 44 & 28.6 & 42 \\
    $P_{19}$ & 250& $m_v^2 c_6^0$ = 100 & 18.5 & 29 & 15.6 & 26 \\
    $P_{20}$ & 250 & $m_v^2 c_7^0$ = 300 & 8.7 & 10 & 17.3 & 28  \\
    $P_{21}$ & 250 & $m_v^2 c_8^0$ = 0.5 & 9.2 & 10 & 14.3 & 24 \\
    $P_{22}$ & 250 & $m_v^2 c_9^0$ = 8 &14.3 & 24 & 19.7 & 31 \\
    $P_{23}$ & 250 & $m_v^2 c_{10}^0$ = 4 & 15 & 25 & 21.6 & 34 \\   
    $P_{24}$ & 250 & $m_v^2 c_{11}^0$ = 0.01 & 15.6 & 26 & 19.1 & 30  \\
    \bottomrule
    \end{tabular}
    \caption{List of the benchmark points studied in this paper. For each benchmark point, we report the dark matter particle mass, and the value of the coupling constant different from zero at the benchmark point. For the Germanium ($j=1$) and Xenon ($j=2$) detectors, we also report $\mu^{(j)}_{\rm tot}(\hat{m}_\chi,\hat{\mathbf{c}})$ and $N_j$, respectively, the expected and the ``observed'' number of scattering events.}
    \label{tab:benchmarks}
\end{table}

\begin{figure}[t]
\begin{center}
\includegraphics[width=\textwidth]{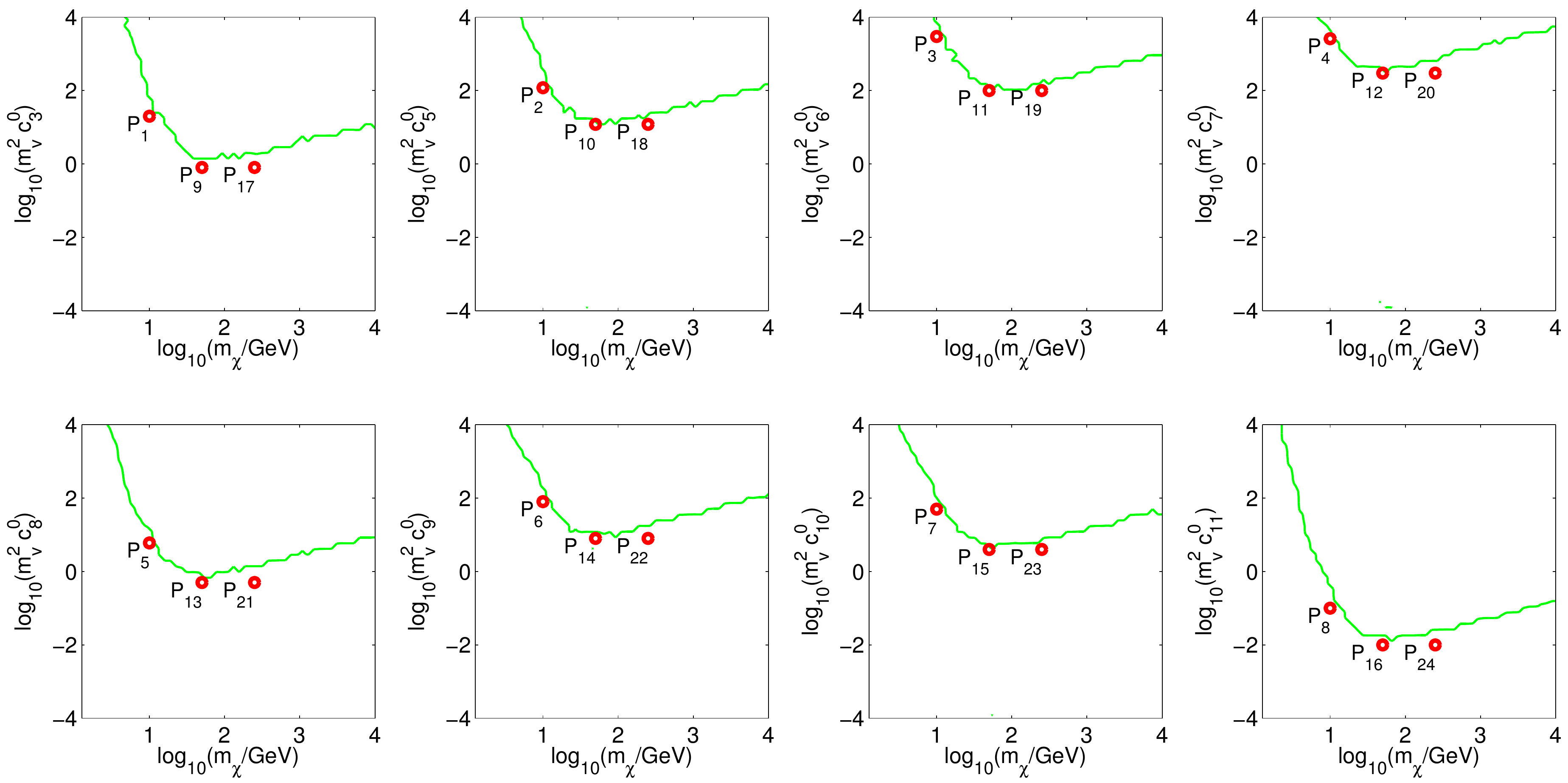}
\end{center}
\caption{Benchmark points studied in this paper represented in the 8 planes $m_\chi$-$c_{i}^0$, with $i~=~3, 5,\dots,11$. At the benchmark points, one of the interaction operators $\mathcal{O}_3$,  $\mathcal{O}_5$,\dots, $\mathcal{O}_{11}$ is responsible for the dark matter-nucleon interaction. Tab.~\ref{tab:benchmarks} shows the properties of the 24 benchmark points. The green contours in the figure represent exclusion limits at the 95\% confidence level derived in Ref.~\cite{Catena:2014uqa} through a global analysis of current direct detection experiments.}
\label{fig:benchmarks}
\end{figure}

\subsection{Synthetic data}
\label{sec:simulations}
Given a detector of type $j$, $j=1,2$, and a benchmark point in the parameter space of the theory defined in Sec.~\ref{sec:theory}, we simulate our synthetic direct detection data as follows. First, we randomly sample the number $N_j$ of observed scattering events from a Poisson distribution of mean $\mu^{(j)}_{\rm tot}(\hat{m}_\chi,\hat{\mathbf{c}})$, where $(\hat{m}_\chi,\hat{\mathbf{c}})$ is the benchmark point. Then, we randomly sample $N_j$ recoil energies, or PE, $\{\hat{E}_j\}_{i=1,\dots,N_j}$ from the spectral function 
\begin{equation}
\hat{f}^{(j)}(\hat{E}_{j}) \equiv f^{(j)}(\hat{E}_j,\hat{m}_\chi,\hat{\mathbf{c}}) 
\end{equation}
where
\begin{equation}
f^{(j)}(\hat{E}_{j},m_\chi,\mathbf{c}) = \frac{1}{\mu^{(j)}_{\rm tot}(m_{\chi},{\mathbf{c}})} \left[ \frac{{\rm d}\mathcal{R}^{(j)}}{{\rm d}\hat{E}_{j}} \left( \hat{E} _{j},m_\chi,\mathbf{c} \right)  + \frac{{\rm d}\mathcal{R}^{(j)}_{\rm B}}{{\rm d}\hat{E}_j} \left( \hat{E} _{j} \right)  \right]\,.
\label{eq:spectralf}
\end{equation}
We apply this procedure to the detectors described in the previous subsection, and to the 24 benchmark points listed in Tab.~\ref{tab:benchmarks}. In the analyses, we repeat the procedure twice for each benchmark point; once for each detector type. From each benchmark point, we therefore produce two datasets $\mathbf{d}_j$, $j=1,2$, one for the Germanium detector and one for the Xenon detector. 

At the benchmark points of Tab.~\ref{tab:benchmarks}, the dark matter particle mass varies from 10 GeV to 250 GeV. The 24 benchmark points in Tab.~\ref{tab:benchmarks} allow to explore the 8 momentum and velocity dependent interaction operators $\mathcal{O}_3, \mathcal{O}_5,\dots\mathcal{O}_{11}$. These benchmark points are compatible with present direct detection experiments, and have been proposed in Ref.~\cite{Catena:2014epa}, studying the prospects for direct detection of dark matter. Fig.~\ref{fig:benchmarks} shows the 24 benchmark points in the 8 planes $m_\chi$-$c_{i}^0$, with $i=3,5,\dots,11$.

\subsection{The likelihood function}
\label{sec:likelihood}
For each benchmark point, we generate two set of synthetic data, $\mathbf{d}_{j}\equiv (N_j,\{\hat{E}_j\}_{i=1,\dots,N_j})$, containing $(N_j+1)$ datapoints. One set of data for the ton-scale Germanium detector ($j=1$) and one for the ton-scale Xenon detector ($j=2$). We assume a Likelihood function 
\begin{eqnarray}
-\ln \mathcal{L}^{(j)}(\mathbf{d}_{j}, |m_\chi,\mathbf{c}) &=& \mu^{(j)}_{\rm tot}(m_\chi,\mathbf{c}) -N_j - N_j \ln [\mu^{(j)}_{\rm tot}(m_\chi,\mathbf{c})/N_j]  \nonumber\\
&+& \sum_{i=1}^{N_j}\log \frac{f^{(j)}(\hat{E}_j,\hat{m}_\chi,\hat{\mathbf{c}})}{f^{(j)}(\hat{E}_{j},m_\chi,\mathbf{c})}\,.
\label{eq:like}
\end{eqnarray}
for each dataset $\mathbf{d}_{j}$, $j=1,2$. The total Likelihood function of the independent datasets $\mathbf{d}_{j}$, $j=1,2$, is therefore given by
\begin{equation}
-\ln \mathcal{L}(\mathbf{d}_1,  \mathbf{d}_2, |m_\chi,\mathbf{c}) =  -\sum_{j=1,2} \ln \mathcal{L}^{(j)}(\mathbf{d}_j, |m_\chi,\mathbf{c}) \,.
\label{eq:like_tot}
\end{equation}
The sum appearing in the second line of Eq.~(\ref{eq:like}) contains the information on the spectrum of observed events, through the dependence on the functions $f^{(j)}$, defined in Eq.~(\ref{eq:spectralf}).

\subsection{Fitting procedures}
\label{sec:plan}
In the analysis of direct detection experiments, it is commonly assumed that:
\begin{enumerate}
\item The dark matter particle interacts with the detector nuclei through the interaction operators $\mathcal{O}_1$ and $\mathcal{O}_4$ only (corresponding to the familiar spin-independent and spin-dependent interactions, respectively).
\item The local population of Milky Way dark matter particles is characterized by a Maxwell-Boltzmann velocity distribution.
\end{enumerate}
Relying on these assumptions, when the actual theory of dark matter violates them, bias the interpretation of direct detection experiments. In order to quantify this bias, we proceed as follows. First, from each benchmark point in Tab.~\ref{tab:benchmarks}, we simulate synthetic direct detection data as explained in Sec.~\ref{sec:simulations}. Then, we analyze the synthetic data using the methods introduced in Sec.~\ref{sec:statistics}. We adopt four fitting procedures: 
\begin{enumerate}
\item {\it Fitting procedure A.} Analyzing the data, we assume as dark matter-nucleon interaction operator the same operator $\mathcal{O}_i$ used in the simulation, and as dark matter velocity distribution our benchmark anisotropic distribution. 
\item {\it Fitting procedure B.} Analyzing the data, we assume $\mathcal{O}_1$ as the correct dark matter-nucleon interaction operator, and our benchmark anisotropic distribution as dark matter velocity distribution. 
\item {\it Fitting procedure C.} Analyzing the data, we assume $\mathcal{O}_4$ as the correct dark matter-nucleon interaction operator, and our benchmark anisotropic distribution as dark matter velocity distribution.  
\item {\it Fitting procedure D.} Analyzing the data, we assume as dark matter-nucleon interaction operator the same operator $\mathcal{O}_i$ used in the simulation, and a Maxwell-Boltzmann distribution with parameters set as in Sec.~\ref{sec:theory} as dark matter velocity distribution. 
\end{enumerate}
Comparing the fitting procedures B, C and D with the fitting procedure A (and the original benchmark points), we estimate the bias in the interpretation of future direct detection experiments potentially induced by incorrect theoretical assumptions. 

\begin{table}
    \centering
    \begin{tabular}{ccccccccc}
    \toprule
    Point   &  $\beta^{\rm A}(m_\chi$) &  $\beta^{\rm A}(f_p)$ &  $\beta^{\rm B}(m_\chi)$ &  $\beta^{\rm B}(f_p)$ &  $\beta^{\rm C}(m_\chi)$ &  $\beta^{\rm C}(f_p)$ &  $\beta^{\rm D}(m_\chi)$&  $\beta^{\rm D}(f_p)$ \\
    \midrule  
$P_{1}$   & 0.06   & 0.01   & 0.23  & 21.11  &  0.22  &  5.26  &  1.05  &  0.73 \\ 
$P_{2}$   & 0.53   & 0.22   & 0.32  & 21.34  &  0.38  &  8.45  &  0.21  &  0.27 \\
$P_{3}$   & 0.17   & 0.24   & 0.55  & 35.84  &  0.54  & 18.55  &  1.04  &  0.70 \\
$P_{4}$   & 0.00   & 0.20   & 0.27  & 21.65  &  0.30  & 11.67  &  0.55  &  0.55 \\
$P_{5}$   & 0.00   & 0.06   & 0.16  & 11.93  &  0.17  &  1.36   & 0.11  &  0.08 \\
$P_{6}$   & 0.31   & 0.08   & 0.45  & 29.46  &  0.48  & 10.22  &  0.73  &  0.87 \\
$P_{7}$   & 0.62   & 0.66   & 0.78  & 23.77  &  0.75  &  7.87   & 0.20  &  0.07 \\
$P_{8}$   & 0.37   & 0.02   & 0.58  &  9.29   & 0.54   & 5.96   & 0.36  &  0.54 \\ 
  \midrule  
$P_{9}$   & 0.02   & 0.25   & 2.39  &  4.96   & 2.51   & 1.34   & -    & - \\
$P_{10}$ &  0.81  &  0.98   & 0.25   &11.09  &  0.29 &   2.62   & - & - \\
$P_{11}$ &   0.07   & 0.29  &  2.49  &  9.83  &  2.57 &   3.34   & -  & - \\
$P_{12}$ &   0.18   & 0.30  &  1.09  & 79.01 &   0.81 &  40.15   & - & - \\
$P_{13}$ &   0.14   & 1.22  &  0.14  & 18.82 &   0.10 &   0.30   & - & - \\
$P_{14}$ &   0.15   & 1.47  &  0.24  & 13.81 &   0.24  &  2.34   & - & - \\
$P_{15}$ &   0.14   & 0.35  &  0.39  &  7.54  &  0.67   & 1.61   & -  & - \\
$P_{16}$ &   0.27   & 1.41  &  0.29  &  2.63  &  0.45   & 3.30   & -  & - \\
  \midrule  
$P_{17}$ &   0.03   & 0.17  &  2.26  &  6.57  &  2.28   & 1.53   & -  & - \\
$P_{18}$ &   0.27   & 0.03  &  2.12  &  9.03  &  2.18   & 1.39   & - & - \\
$P_{19}$ &   0.12   & 0.13  &  2.25  & 11.95 &   2.27  &  4.06  &  - & - \\
$P_{20}$ &   0.25   & 0.40  &  0.37  & 10.74 &   0.25  &  5.32  &  - & - \\
$P_{21}$ &   0.41   & 0.16  &  0.33  &  5.87  &  0.17   & 0.42   & -  & - \\
$P_{22}$ &   0.17   & 0.06  &  1.65  &  7.04  &  1.72   & 0.96   & -  & - \\
$P_{23}$ &   0.34   & 0.06  &  1.75  & 6.74   & 1.81   & 0.32  &  -  & - \\ 
$P_{24}$ &   0.37   & 0.08  &  1.71  &  1.23  &  1.76  &  5.22  &  -  & - \\
  \bottomrule
    \end{tabular}
    \caption{List of bias parameters. For each benchmark point, we report the bias parameters of $\log_{10}(m_\chi/{\rm GeV})$ and $\log_{10}(f_p)$ resulting from four different analyses. We denote by $b^k(x)$ the bias parameter of $\log_{10}(x)$, as obtained from the fitting procedure $k$, with $k={\rm A}, {\rm B}, {\rm C}, {\rm D}$. Sec.~\ref{sec:plan} describes the four different fitting procedures.}
    \label{tab:beta}
\end{table}

\begin{figure}[t]
\begin{center}
\includegraphics[width=\textwidth]{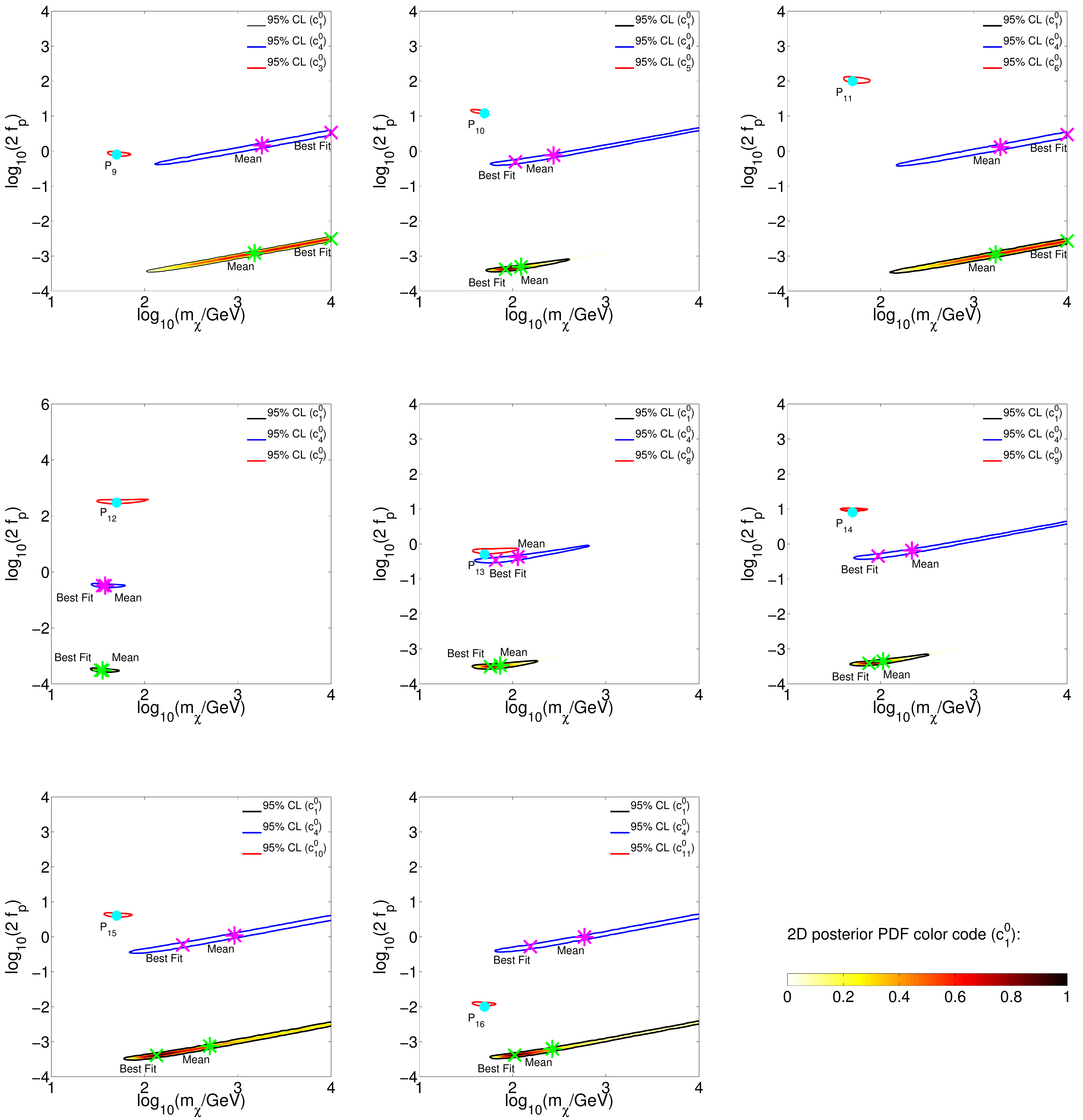}
\end{center}
\caption{Analysis of the benchmark points $P_{9}$ -- $P_{16}$. The eight panels show the 2D 95\% confidence intervals that we obtain fitting $m_\chi$ and $c^0_1$ (black contours), $m_\chi$ and $c^0_4$ (blue contours), as well as $m_\chi$ and the constant $c_i^0\neq 0$ in Tab.~\ref{tab:benchmarks} (red contours) to the synthetic data. Green (magenta) crosses and stars represent the best fit values and the posterior means associated with the black (blue) contours. Cyan dots denote the benchmark points. In each panel, we report the posterior PDF that we find fitting $m_\chi$ and $c^0_1$ to the synthetic data. Fitting the interaction operators $\mathcal{O}_1$ and $\mathcal{O}_4$ to future direct detection data, when the interaction underlying the data is of a different type, induces a bias in the interpretation the experimental results. There are cases in which this bias cannot be identified through a simple goodness of fit test (see text around Fig.~\ref{fig:P11P13}).}
\label{fig:50}
\end{figure}
\begin{figure}[t]
\begin{center}
\begin{minipage}[t]{0.49\linewidth}
\centering
\includegraphics[width=\textwidth]{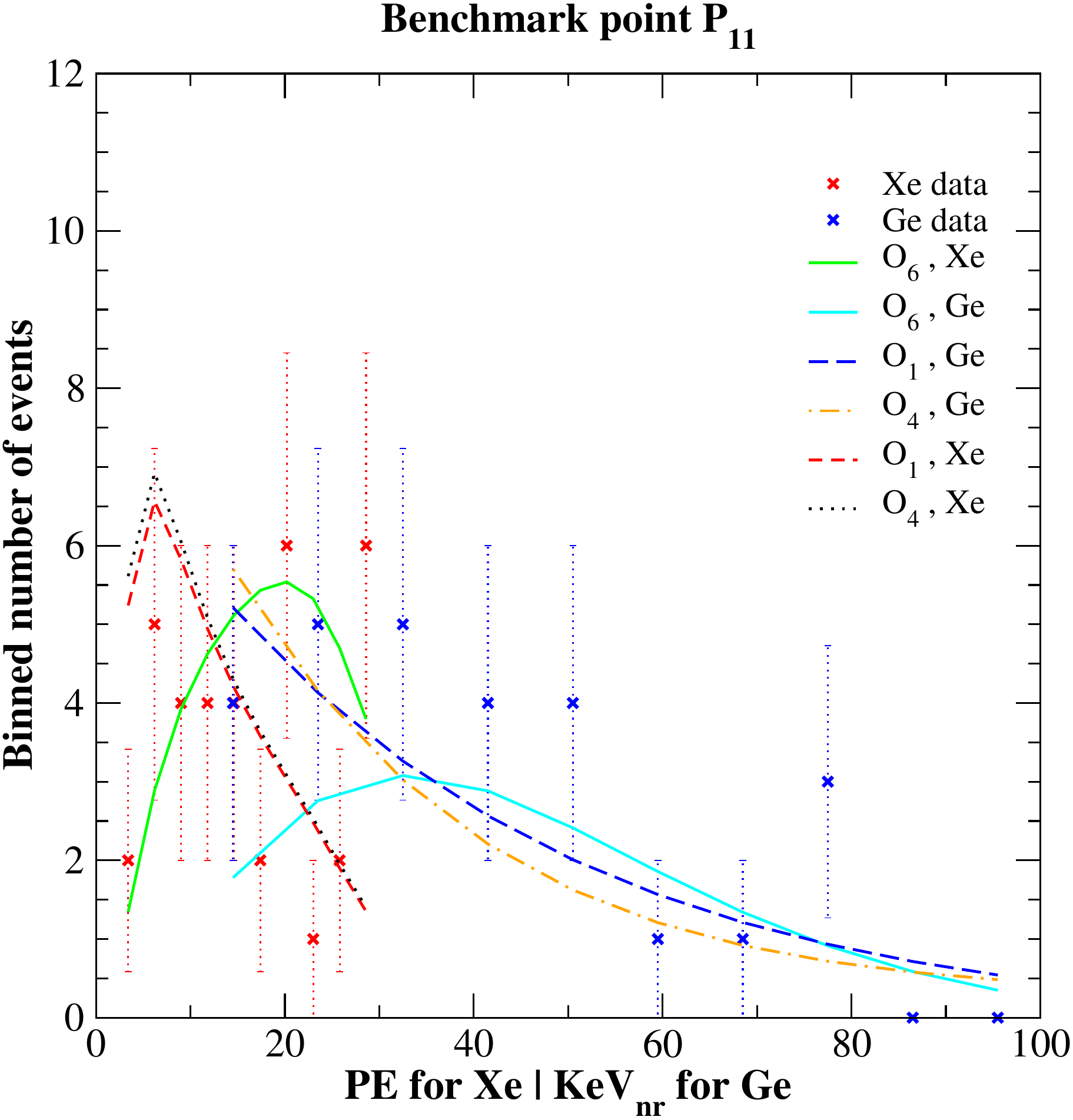}
\end{minipage}
\begin{minipage}[t]{0.49\linewidth}
\centering
\includegraphics[width=\textwidth]{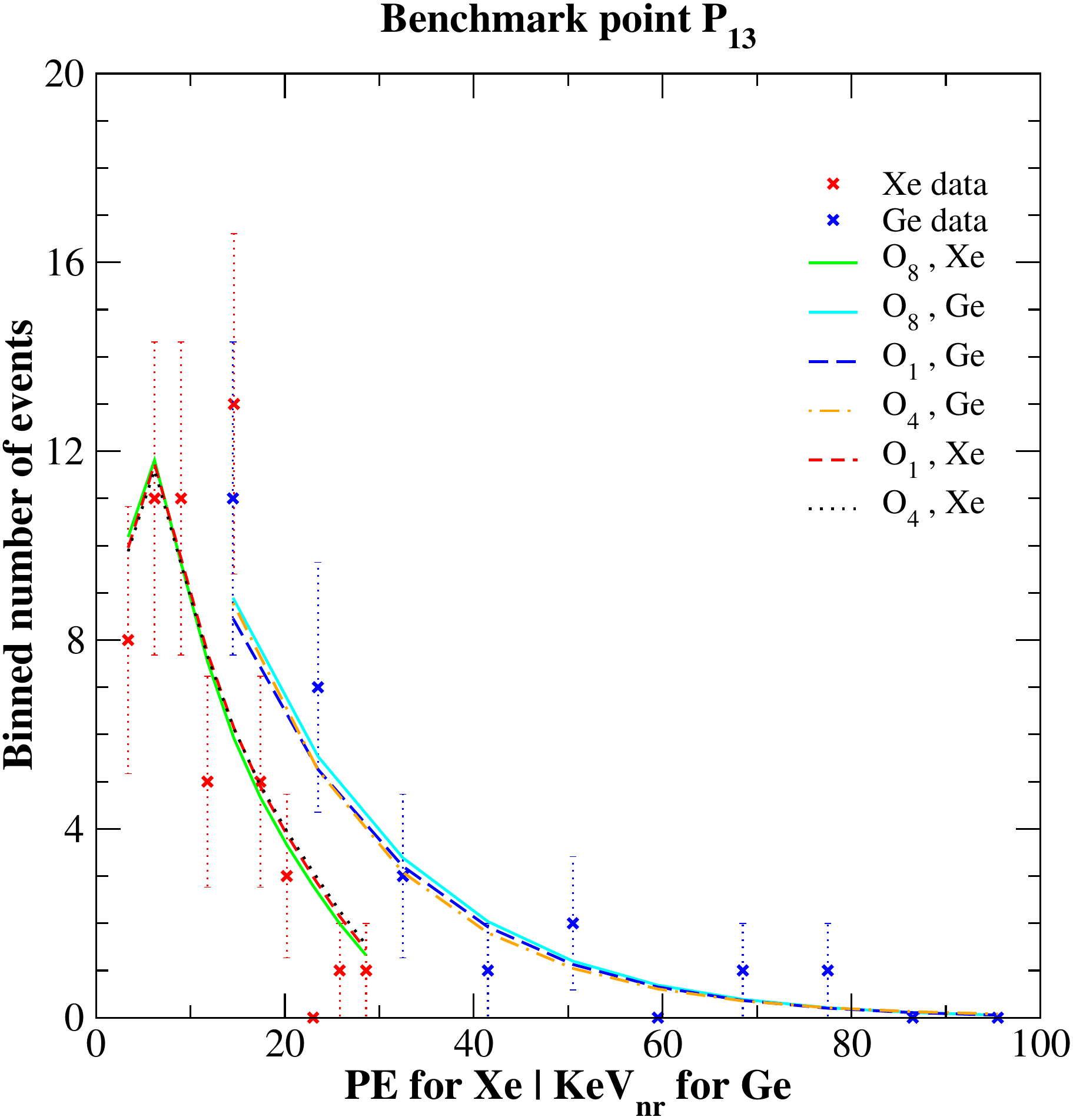}
\end{minipage}
\end{center}
\caption{Synthetic data generated from the benchmark points $P_{11}$ (left panel) and $P_{13}$ (right panel). We bin the events measured by the Germanium detector in 10 energy bins, and the events measured by the Xenon detector in 10 PE bins. We compare the binned data to the expected number of events at each bin and for each detector. The expected number of events shown in the figures has been obtained integrating Eqs.~(\ref{eq:rate-Ge}) and (\ref{eq:rate_Xe}). We set the model parameters at the best fit points found using the unbinned likelihood (\ref{eq:like_tot}), and the fitting procedures A, B and C. The fitting procedures A, B and C assume that at $P_{11}$ ($P_{13}$) dark matter interacts with the detector nuclei through the interaction operators $\mathcal{O}_6$ ($\mathcal{O}_8$), $\mathcal{O}_1$ and $\mathcal{O}_4$, respectively. To help the comparison, we have connected points representing different theoretical predictions with colored lines. The notation is the one explained in the legends. For each bin, error bars corresponds to $\pm$ the square root of the observed number of events.}
\label{fig:P11P13}
\end{figure}
\begin{figure}[t]
\begin{center}
\includegraphics[width=\textwidth]{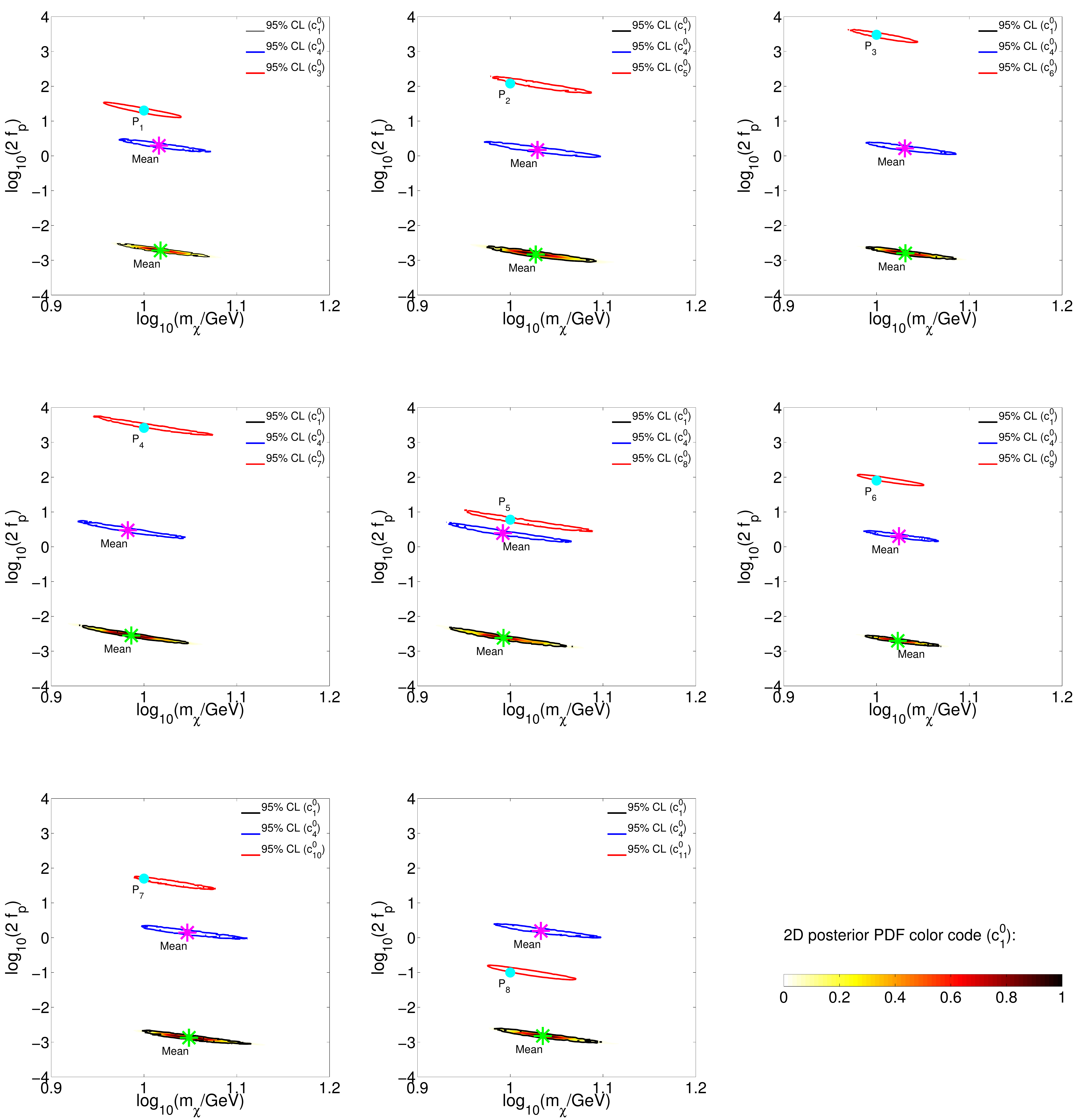}
\end{center}
\caption{Same as for Fig.~\ref{fig:50}, but for the benchmark points $P_1$ -- $P_8$. In the panels, we do not report the best fit points, since they are essentially superimposed to the posterior means.}
\label{fig:10}
\end{figure}

\begin{figure}[t]
\begin{center}
\includegraphics[width=\textwidth]{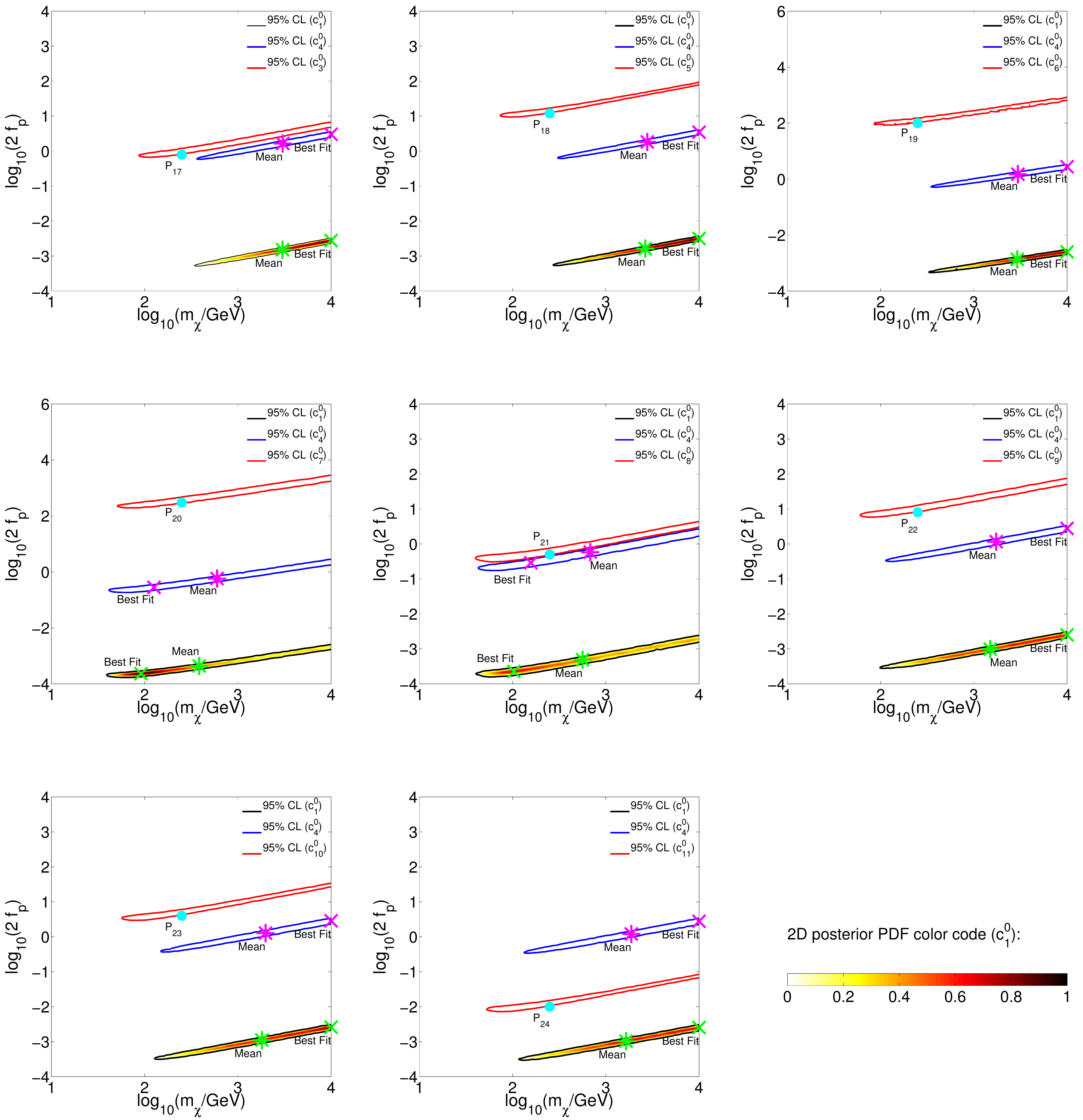}
\end{center}
\caption{Same as for Fig.~\ref{fig:50}, but for the benchmark points $P_{17}$ -- $P_{24}$.}
\label{fig:250}
\end{figure}

\section{Results}
\label{sec:results}
In this section we analyze our synthetic direct detection data following the four fitting procedures described in Sec.~\ref{sec:plan}. We start with an analysis of the bias induced by assuming $\mathcal{O}_1$ and $\mathcal{O}_4$ as the only dark matter-nucleon interaction operators in the determination of $m_\chi$ and $\mathbf{c}$ (Sec.~\ref{sec:results1}). Then, we analyze the bias induced by the Maxwell-Boltzmann approximation to our benchmark anisotropic velocity distribution in the interpretation of future direct detection experiments (Sec.~\ref{sec:results2}). In the figures, for all the interaction operators we denote by $f_p=m_v^2 c_i^0/2$, $i=3, 5, \dots,11$, the dimensionless dark matter-nucleon coupling constant, in the limit of isoscalar interactions. 

\subsection{Data analysis with a particle physics bias}
\label{sec:results1}
In this section, we analyze the synthetic data generated from the benchmark points $P_1$ -- $P_{24}$ adopting the fitting procedures A, B and C defined in Sec.~\ref{sec:plan}. 

We start with a detailed analysis of the benchmark points $P_9$ -- $P_{16}$. We report the results of this analysis in Fig.~\ref{fig:50}. The benchmark points $P_9$ -- $P_{16}$ are characterized by a dark matter particle mass $m_\chi=50$~GeV. Their properties are summarized in Tab.~\ref{tab:benchmarks}. Different panels in Fig.~\ref{fig:50} correspond to distinct benchmark points. In the panels, the benchmark points are represented by cyan dots. For each benchmark point, we present our results in terms of 2D 95\% confidence intervals in the $m_\chi$-$f_p$ plane. Red contours, black contours and blue contours correspond to the 2D 95\% confidence intervals that we obtain applying, respectively, the fitting procedures A, B and C to the benchmark points $P_9$ -- $P_{16}$. In order to compare the Bayesian and the frequentist approach, in each panel we report the 2D posterior PDF that we derive analyzing the synthetic data with the fitting procedure B. Finally, green (magenta) crosses and stars in the panels identify the best fit points and the posterior means resulting from the fitting procedure B (C). 

First, we comment on the results that we obtain applying the fitting procedure A to the synthetic data generated from the benchmark points $P_9$ -- $P_{16}$. In this analysis, we find that the bias parameter of $\log_{10}(m_\chi/{\rm GeV})$ is always less than 1, ranging from $\beta~=~0.02$ at $P_{9}$ to $\beta=0.81$ at $P_{10}$. The bias parameter of $\log_{10}(f_p)$ is larger and it varies from $\beta=0.25$ at $P_{9}$ to $\beta=1.47$ at $P_{14}$ (see Tab.~\ref{tab:beta}). We conclude that the fitting procedure A leads to a good reconstruction of the benchmark points $P_9$ -- $P_{16}$.

We now apply the fitting procedure B to the synthetic data generated from the benchmark points $P_9$ -- $P_{16}$. The fitting procedure B allows us to study the bias induced by assuming $\mathcal{O}_1$ as the correct dark matter nucleon interaction operator in the determination of the dark matter particle properties. Interestingly, the benchmark points $P_9$ -- $P_{16}$ can be divided in two groups, depending on the accuracy within which $m_\chi$ is determined in this analysis. 

A first group includes the benchmark points $P_{9}$, $P_{11}$, $P_{15}$ and $P_{16}$. At these benchmark points, the confidence intervals around the best fit values of $m_\chi$ are large, if compared with the confidence intervals found with the fitting procedure A. In order to further investigate this difference between the fitting procedures A and B, we perform a chi-square goodness of fit test, using the best fit points found within the two approaches. First, we bin the recoil energies observed by the Germanium detector in 10 energy bins, and the PE observed by the Xenon detector in 10 PE bins. Then, we calculate the expected number of events in the 10 energy bins, and in the 10 PE bins, setting the model parameters at the best fit points found with the fitting procedures A and B. Combining the binned data with the theoretical expectations, we construct a reduced chi-square $\chi_{\rm red}^2$ with 18 degrees of freedom for the fitting procedures A and B. Using the synthetic data generated from the benchmark point $P_{11}$, for instance, we find $\chi_{\rm red}^2=1.34$ and $\chi_{\rm red}^2=1.88$ for the fitting procedures A and B, respectively. We obtain similar results for the benchmark points $P_{9}$, $P_{15}$ and $P_{16}$. This study indicates that for $m_\chi=50$~GeV, the operator $\mathcal{O}_1 = 1_{\chi} 1_{N}$ cannot provide a good fit to the synthetic data generated from the interaction operators $\mathcal{O}_3 = -i\vec{S}_N\cdot(\vec{q}/m_N\times\vec{v}^{\perp}_{\chi N})$, $\mathcal{O}_6 = (\vec{S}_\chi\cdot\vec{q}/m_N)(\vec{S}_N\cdot\vec{q}/m_N)$, $\mathcal{O}_{10} = -i\vec{S}_N\cdot\vec{q}m_N$ and $\mathcal{O}_{11} = -i\vec{S}_\chi\cdot\vec{q}/m_N$. Here, we use the same notation introduced in the caption of Tab.~\ref{tab:operators}. Fig.~\ref{fig:P11P13} (left panel) shows the binned data, and the expected number of events in each bin for the benchmark point $P_{11}$.

A second group includes the benchmark points $P_{10}$, $P_{12}$, $P_{13}$ and $P_{14}$. At these benchmark points, we find confidence intervals based on the fitting procedure B, comparable with those that we obtain using the fitting procedure A. A chi-square goodness of fit test applied to the synthetic data generated from these benchmark points gives as a result similar values of $\chi_{\rm red}^2$ for the fitting procedures A and B. For instance, at the benchmark point $P_{13}$, we find $\chi_{\rm red}^2=1.13$ and $\chi_{\rm red}^2=1.15$ for the fitting procedures A and B, respectively. We obtain comparable values of $\chi_{\rm red}^2$ for the benchmark points $P_{10}$, $P_{12}$ and $P_{14}$. Fig.~\ref{fig:P11P13} (right panel) reports the binned data, and the theoretical expectations for the benchmark point $P_{13}$. In summary, for $m_\chi=50$~GeV the operator $\mathcal{O}_1$ provides a good fit to the synthetic data generated from the interaction operators $\mathcal{O}_5 = -i\vec{S}_\chi\cdot(\vec{q}/m_N\times\vec{v}^{\perp}_{\chi N})$, $\mathcal{O}_7 = \vec{S}_{N}\cdot \vec{v}^{\perp}_{\chi N}$, $\mathcal{O}_8 = \vec{S}_{\chi}\cdot \vec{v}^{\perp}_{\chi N}$, and $\mathcal{O}_9 = -i\vec{S}_\chi\cdot(\vec{S}_N\times\vec{q}/m_N)$. The interaction operator $\mathcal{O}_1$ provides a good fit to these data for the following reason. For $m_{\chi}=50$~GeV, features in the energy recoil spectrum depending on the velocity or on the momentum transfer, which could distinguish the operators $\mathcal{O}_5$, $\mathcal{O}_7$, $\mathcal{O}_8$ and $\mathcal{O}_9$ from the operator $\mathcal{O}_1$, move below the experimental thresholds (see Fig.~\ref{fig:P11P13}, right panel).

We now comment on the bias parameters resulting from the fitting procedure B applied to the benchmark points $P_9$ --$P_{16}$. At these benchmark points, the bias parameter of $\log_{10}(m_\chi/{\rm GeV})$ ranges from $\beta=0.14$ at $P_{13}$ to $\beta=2.49$ at $P_{11}$ (see Tab.~\ref{tab:beta}). Therefore, the bias resulting from the fitting procedure B is generically larger than the bias found with the fitting procedure A. 

We interpret these large values of $\beta$ as follows. Let us consider for a moment the realist case in which the correct dark matter-nucleon interaction is unknown. In presence of a candidate dark matter signal, a reasonable approach to data analysis consists in fitting the operators $\mathcal{O}_i$, $i=1,3,\dots, 11$ to the data independently. A posteriori, a goodness of fit test would identify the correct dark matter-nucleon interaction operator: one would interpret as the correct dark matter-nucleon interaction operator, the operator leading to the value of $\chi_{\rm red}^2$ unambiguously closer to one. From this reasoning, we conclude that a large value of $\beta$ alone does not necessarily indicate the presence of a real bias in the interpretation of the data. Only when $\beta$ is large, say larger than~$\sim1$, and the values of $\chi_{\rm red}^2$ resulting from the fitting procedures A and B are comparable, there is a significant bias in the results, which cannot be identified from the data directly. 

Interestingly, at the benchmark point $P_{12}$, where dark matter interacts with the detector nuclei through the operator $\mathcal{O}_7 = \vec{S}_{N}\cdot \vec{v}^{\perp}_{\chi N}$, the bias parameter of $\log_{10}(m_\chi/{\rm GeV})$ is equal to $1.09$, and the bias parameter of $\log_{10}(f_p)$ is equal to $79.01$. At this benchmark point, from a chi-square goodness of fit test we find $\chi_{\rm red}^2=1.15$ and $\chi_{\rm red}^2=1.38$, binning the synthetic data and using the best fit points of the fitting procedures A and B, respectively. 

At the benchmark points $P_{9}$ -- $P_{16}$, the bias parameter of $\log_{10}(f_p)$ is always much larger than 1, with the exception of the benchmark point $P_{16}$, where $\beta=2.63$ (see Tab.~\ref{tab:beta}). A bias in $\log_{10}(f_p)$ is clearly related to the fact that interaction operators other than $\mathcal{O}_1$ and $\mathcal{O}_4$ are momentum and velocity suppressed. Hence, a large value of $f_p$ is required in order to compensate for this suppression (see below, for an exception in the case of $\mathcal{O}_4$). 

The fitting procedure C applied to the synthetic data generated from the benchmark points $P_9$ -- $P_{16}$ allows us to study the bias induced by assuming $\mathcal{O}_4$ as the correct dark matter-nucleon interaction operator in the determination of $m_\chi$ and $f_p$. The results of this analysis are similar to those previously discussed for the case of the interaction operator $\mathcal{O}_1$. The only significant difference between the two cases is in the best fit value of the dark matter-nucleon coupling constant $f_p$ at the benchmark point $P_{16}$. At $P_{16}$, the best fit value of $f_p$ is larger than its benchmark value, contrary to what we observe at the other benchmark points. The reason is the following. At the benchmark point $P_{16}$, the dark matter-nucleon interaction is described by the operator $O_{11}$. This interaction operator generates the nuclear response function $W_{M}^{\tau\tau'}$ in the dark matter-nucleus scattering. $W_{M}^{\tau\tau'}$ is significantly larger than the nuclear response function associated with the operator $\mathcal{O}_4$, for Germanium and Xenon targets~\cite{Fitzpatrick:2012ix}. Therefore, a large value of $f_p=m_v^2 c_4^0/2$ compensates for the lower nuclear response function generated by $\mathcal{O}_4$. Figs.~\ref{fig:50} and \ref{fig:P11P13} show, respectively, the 95\% 2D confidence intervals and the results of a chi-square goodness of fit analysis based on the fitting procedure C. Tab.~\ref{tab:beta} contains the values of the bias parameters of $\log_{10}(m_\chi/{\rm GeV})$ and $\log_{10}(f_p)$ extracted from the fitting procedure C. 

\begin{figure}[t]
\begin{center}
\includegraphics[width=\textwidth]{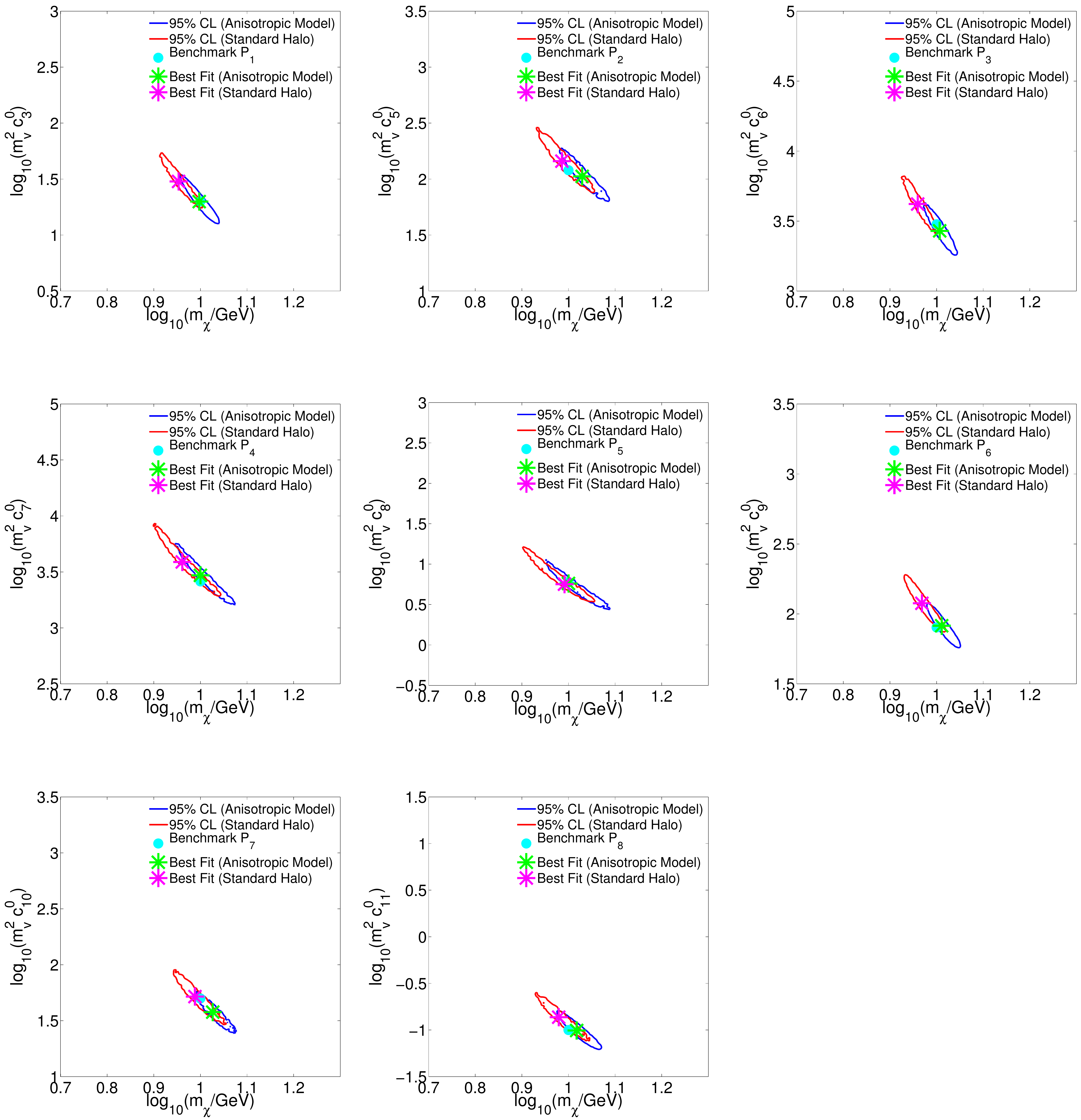}
\end{center}
\caption{Study of the benchmark points $P_{1}$ -- $P_{8}$ in a compared analysis of the fitting procedures A and D (see Sec.~\ref{sec:plan} for a definition of the two fitting procedures). The former approach assumes an anisotropic dark matter velocity distribution in the interpretation of the synthetic data, the latter the standard dark matter halo approximation based on a Maxwell-Boltzmann distribution. Blue contours and red contours correspond to the 2D 95\% confidence intervals that we obtain from the fitting procedures A and D, respectively. Stars denote the best fit points, whereas dots correspond to the benchmark points.}
\label{fig:astro}
\end{figure}

We conclude this subsection with an analysis of the remaining benchmark points. Figs.~\ref{fig:10} and \ref{fig:250} show the 2D 95\% confidence intervals that we obtain from an analysis of the synthetic data generated from the benchmark points $P_1$ -- $P_8$, and $P_{17}$ -- $P_{24}$, respectively. The benchmark points $P_1$ -- $P_8$ are characterized by $m_\chi=10$~GeV, whereas at the benchmark points $P_{17}$ -- $P_{24}$ $m_\chi=250$~GeV. For the benchmark points $P_1$ -- $P_8$, we find a small bias induced by the fitting procedures B and C in the determination of the dark matter particle mass. The corresponding bias parameter $\beta$ varies from $\sim 0.15$, in the case of the benchmark point $P_{5}$ to~$\sim 0.75$, in the case of the benchmark point $P_{7}$, independently of whether we fit the operator $\mathcal{O}_1$ or the operator $\mathcal{O}_4$ to the synthetic data. This small bias is an expected result, since for light dark matter candidates features in the recoil energy spectrum distinguishing a momentum dependent operator from $\mathcal{O}_1$ and $\mathcal{O}_4$ move below the experimental thresholds. Finally, at the benchmark points $P_{17}$ -- $P_{24}$, we reconstruct the dark matter particle mass with large errors in all cases. In addition, the best fit values of $\log_{10}(m_\chi/{\rm GeV})$ move towards very large dark matter masses, when we analyze the synthetic data adopting the fitting procedures B and C, with the exception of the benchmark points $P_{12}$ and $P_{13}$. Tab.~\ref{tab:beta} shows a list of bias parameters for the benchmark points $P_1$ -- $P_8$, and $P_{17}$ -- $P_{24}$.

\subsection{Data analysis with an astrophysical bias}
\label{sec:results2}
We conclude Sec.~\ref{sec:results} studying the benchmark points $P_{1}$ -- $P_{8}$ in a compared analysis of the fitting procedures A and D. Our aim is to quantify the bias induced by the Maxwell-Boltzmann approximation to our benchmark anisotropic velocity distribution in the determination of the dark matter particle mass and coupling constants. 

We focus on the benchmark points $P_{1}$ -- $P_{8}$, since at these points the dark matter particle is light. A light dark matter particle can transfer a detectable amount of energy to a target nucleus, only if it has a sufficiently high speed. Since the Maxwell-Boltzmann distribution mostly differs from our benchmark distribution in the high velocity tail (see Fig.~7 in Ref.~\cite{Bozorgnia:2013pua}), we expect that the benchmark points $P_{1}$ -- $P_{8}$ are the most affected by ``an astrophysical bias'' of the type mentioned above.

Fig.~\ref{fig:astro} shows the 2D 95\% confidence intervals that we obtain applying the fitting procedures A (blue contours) and D (red contours) to the synthetic data generated from the benchmark points $P_{1}$ -- $P_{8}$. Green and magenta stars represent the best fit points of the fitting procedures A and D, respectively. Distinct panels refer to different benchmark points. In all panels, the best fit values of $m_\chi$ that we obtain with the fitting procedure A are larger, compared to those found with the fitting procedure D. The reason for this general trend is the following. The anisotropic velocity distribution of Ref.~\cite{Bozorgnia:2013pua} predicts less dark matter particles with high speed in the galactic halo, as compared to the Maxwell Boltzmann approximation (again, see Fig.~7 in Ref.~\cite{Bozorgnia:2013pua}). Hence, in order to produce a signal above the experimental thresholds, a larger value of the dark matter particle mass is necessary, when using the fitting procedure A.

The bias parameter of $\log_{10}(m_\chi/{\rm GeV})$ that we obtain from the fitting procedure D ranges from 0.11 at $P_{5}$, to 1.05 at $P_{1}$. This result is in agreement with the conclusions of Ref.~\cite{Pato:2012fw}, where studying the interaction operator $\mathcal{O}_1$, the authors find a relative error in the reconstruction of the dark matter particle mass of the order of 1 standard deviation, for certain astrophysical configurations, and with a slightly different definition of bias. Similarly, the bias parameter of $\log_{10}(f_p)$ varies from $0.07$ at $P_{7}$, to $0.87$ at $P_{6}$.  

\begin{figure}[t]
\begin{center}
\begin{minipage}[t]{0.49\linewidth}
\centering
\includegraphics[width=\textwidth]{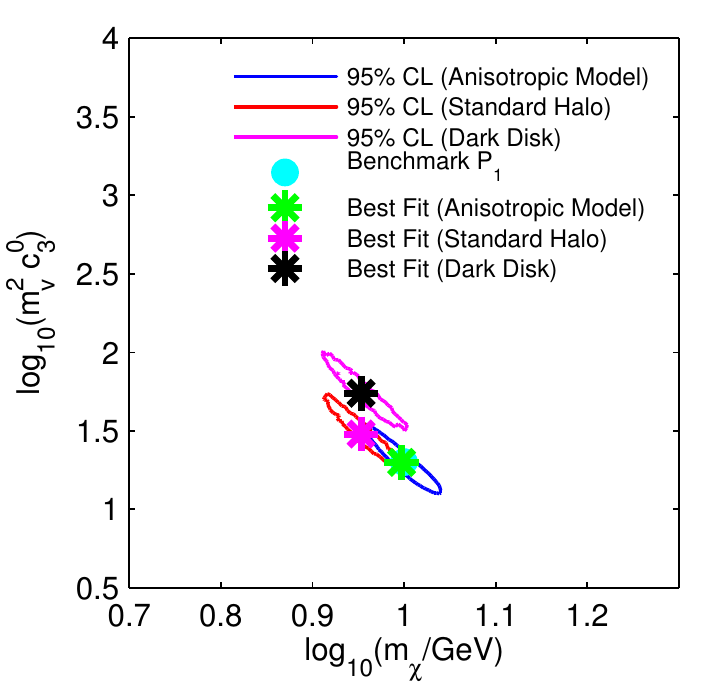}
\end{minipage}
\begin{minipage}[t]{0.49\linewidth}
\centering
\includegraphics[width=0.98\textwidth]{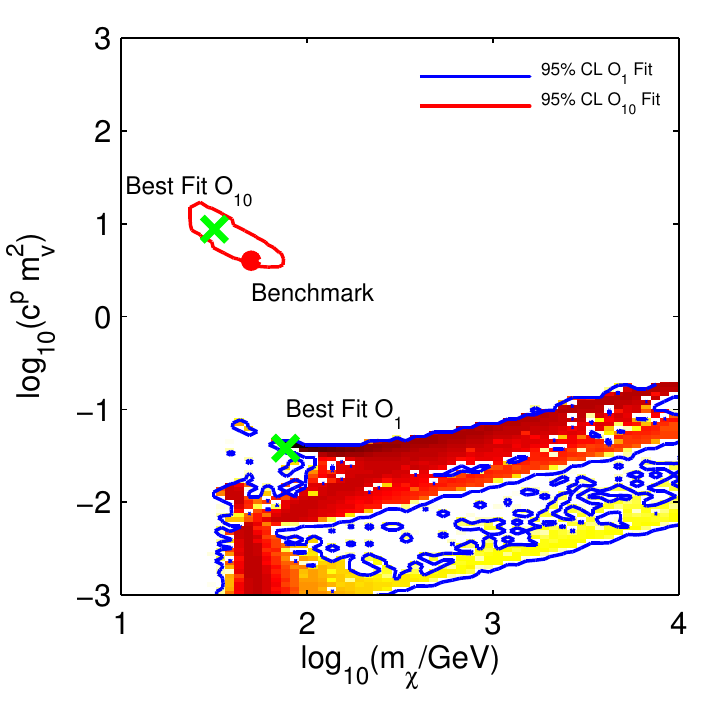}
\end{minipage}
\end{center}
\caption{{\it Left panel.} The blue and red contours have the same meaning as in the top-left panel of Fig.~\ref{fig:astro}. The magenta contour represents the 2D 95\% confidence interval obtained by analyzing the synthetic data generated from $P_1$ assuming in the fit a dark matter disk with parameters set as in Fig.~21 of Ref.~\cite{Peter:2013aha}. {\it Right panel.} Analysis of a benchmark point with $m_\chi=50$~GeV, and $m_v^2 c_{10}^0=m_v^2 c_{10}^1=4$ (not included in Tab.~\ref{tab:benchmarks}). From this benchmark point, we simulate the data of a ton-scale version of the COUPP experiment (modeled as in Ref.~\cite{Catena:2014uqa}, with exposure multiplied by a factor 100). We then fit these data varying $c^0_{10}$ and $c^1_{10}$ only (red contour), or $c^0_{1}$ and $c^1_{1}$ only (blue contour). In the case of the constants $c^0_{1}$ and $c^1_{1}$, we also show the 2D likelihood in the $c^p$--$m_\chi$ plane, using the same color code adopted in the previous figures for the 2D posterior PDF. $c^p$ is the dark matter coupling to protons in Eq.~(\ref{eq:couplings}) (we have omitted the $i$ index).}
\label{fig:mix}
\end{figure}

Comparing the results of this section with those of Sec.~\ref{sec:results1}, we conclude that at present the most serious source of theoretical bias in dark matter direct detection is our poor knowledge of the dark matter-nucleon interaction. This conclusion remains true even when velocity distributions significantly different from the Maxwellian distribution are considered, as it is shown in the left panel of Fig.~\ref{fig:mix}.  

\section{Conclusions}
\label{sec:conclusions}
Incorrect assumptions about the dark matter-nucleon interaction, or the local dark matter velocity distribution will bias the interpretation of future direct detection experiments. We have performed a quantitative study of the theoretical bias in dark matter direct detection, focusing on the assumption of momentum/velocity independent dark matter-nucleon interactions, and on the Maxwell-Boltzmann approximation.

We have addressed this problem within the effective theory of isoscalar dark matter-nucleon interactions mediated by a heavy spin-1 or spin-0 particle. From the 8 momentum and velocity dependent interaction operators predicted by the theory, we have simulated the data of future ton-scale Germanium and Xenon detectors, assuming an anisotropic dark matter velocity distribution. Interpreting the simulated data, we have instead made different assumptions, thereby introducing a theoretical bias in the analysis. We have either assumed a momentum and velocity independent dark matter-nucleon interaction, or a Maxwell-Boltzmann dark matter velocity distribution. Comparing the best fit points with the original benchmark points, we have hence estimated the theoretical bias in dark matter direct detection, using frequentist and Bayesian statistical methods. 

We have found examples where the operator $\mathcal{O}_i$, $i\neq1,4$ generating the observed signal, the operator $\mathcal{O}_1$ and $\mathcal{O}_{4}$ fit the synthetic data equally well, in a chi-square goodness of fit test. However, fitting $\mathcal{O}_1$ and $\mathcal{O}_{4}$ to the synthetic data, we find a biased reconstruction of $m_\chi$ and $f_p$.

For instance, we have found that the signal produced in a ton-scale detector by a dark matter particle of mass $50$~GeV, coupling to the nucleons through the operator $\mathcal{O}_7~=~\vec{S}_{N}\cdot \vec{v}^{\perp}_{\chi N}$ with a strength $f_p=300$, can be confused with the signal produced by a dark matter particle of mass $33.8$~GeV, coupling to the nucleons through the operator $\mathcal{O}_1~=~1_{\chi} 1_{N}$ with a strength $f_p=3\times 10^{-4}$ (see Fig.~\ref{fig:50}, central-left panel). In this example, the familiar spin-independent interaction (i.e. $\mathcal{O}_1$) provides a good fit to the synthetic data generated from $\mathcal{O}_7$, producing however significantly biased results. 

In the last part of the paper, we have found that for a 10~GeV dark matter candidate, incorrect astrophysical assumptions induce an error in the reconstruction of $m_\chi$ and $f_p$ of the order of 1 standard deviation. 

Possible extensions of the present study include an analysis of the isovector couplings, the exploration of larger ensembles of ton-scale detectors, and an investigation of the modulation signal as an additional tool to possibly discriminate the different dark matter-nucleon interaction types. We leave these analyses for future work. The results of a preliminary study of the isovector coupling $c^1_{10}$ that assumes a ton-scale version of the COUPP detector is reported in the right panel of Fig.~\ref{fig:mix}.

We conclude that a potentially important source of theoretical bias inheres within common assumptions in the field of dark matter direct detection. Extracting the correct dark matter-nucleon interaction from a multi-dimensional analysis of the data directly will be important to fully exploit the next generation of direct detection data~\cite{Catena:2014epa,Gluscevic:2014vga}. Marginalizing over (in a Bayesian approach) or profiling out (in a frequentist approach) the astrophysical uncertainties pertaining the dark matter direct detection could mitigate the astrophysical bias identified in this study~\cite{Pato:2012fw,Peter:2013aha}.  

\acknowledgments R.C. acknowledges partial support from the European Union FP7 ITN INVISIBLES (Marie Curie Actions, PITN-GA-2011-289442).

\appendix
\section{Dark matter response functions}
\label{sec:dmrfun}
In the following, we list the dark matter response functions appearing in Eq.~(\ref{Ptot}). In the calculations, we assume for definiteness that the dark matter particle has spin $j_\chi=1/2$.

\allowdisplaybreaks
\begin{eqnarray}
\label{eq:Rfunctions}
 R_{M}^{\tau \tau^\prime}({v}^{\perp 2}_{\chi T}, {{q}^{2} \over m_N^2}) &=& c_1^\tau c_1^{\tau^\prime } + {j_\chi (j_\chi+1) \over 3} \left[ {{q}^{2} \over m_N^2} {v}^{\perp 2}_{\chi T} c_5^\tau c_5^{\tau^\prime }+{v}^{\perp 2}_{\chi T} c_8^\tau c_8^{\tau^\prime }
+ {{q}^{2} \over m_N^2} c_{11}^\tau c_{11}^{\tau^\prime } \right] \nonumber \\
 R_{\Phi^{\prime \prime}}^{\tau \tau^\prime}({v}^{\perp 2}_{\chi T}, {{q}^{2} \over m_N^2}) &=& {{q}^{2} \over 4 m_N^2} c_3^\tau c_3^{\tau^\prime } 
 \nonumber \\
 R_{\Phi^{\prime \prime} M}^{\tau \tau^\prime}({v}^{\perp 2}_{\chi T}, {{q}^{2} \over m_N^2}) &=&  c_3^\tau c_1^{\tau^\prime } 
 \nonumber \\
  R_{\tilde{\Phi}^\prime}^{\tau \tau^\prime}({v}^{\perp 2}_{\chi T}, {{q}^{2} \over m_N^2}) &=& 0
  \nonumber \\
   R_{\Sigma^{\prime \prime}}^{\tau \tau^\prime}({v}^{\perp 2}_{\chi T}, {{q}^{2} \over m_N^2})  &=&{{q}^{2} \over 4 m_N^2} c_{10}^\tau  c_{10}^{\tau^\prime } +
  {j_\chi (j_\chi+1) \over 12} \left[ c_4^\tau c_4^{\tau^\prime} + 
 {{q}^{2} \over m_N^2} ( c_4^\tau c_6^{\tau^\prime }+c_6^\tau c_4^{\tau^\prime })+
 {{q}^{4} \over m_N^4} c_{6}^\tau c_{6}^{\tau^\prime } \right] \nonumber \\
    R_{\Sigma^\prime}^{\tau \tau^\prime}({v}^{\perp 2}_{\chi T}, {{q}^{2} \over m_N^2})  &=&{1 \over 8} \left[ {{q}^{2} \over  m_N^2}  {v}^{\perp 2}_{\chi T} c_{3}^\tau  c_{3}^{\tau^\prime } + {v}^{\perp 2}_{\chi T}  c_{7}^\tau  c_{7}^{\tau^\prime }  \right]
       + {j_\chi (j_\chi+1) \over 12} \left[ c_4^\tau c_4^{\tau^\prime} + 
        {{q}^{2} \over m_N^2} c_9^\tau c_9^{\tau^\prime }
        \right] \nonumber \\
     R_{\Delta}^{\tau \tau^\prime}({v}^{\perp 2}_{\chi T}, {{q}^{2} \over m_N^2})&=&  {j_\chi (j_\chi+1) \over 3} \left[ {{q}^{\,2} \over m_N^2} c_{5}^\tau c_{5}^{\tau^\prime }+ c_{8}^\tau c_{8}^{\tau^\prime } \right] \nonumber \\
 R_{\Delta \Sigma^\prime}^{\tau \tau^\prime}({v}^{\perp 2}_{\chi T}, {{q}^{2} \over m_N^2})&=& {j_\chi (j_\chi+1) \over 3} \left[c_{5}^\tau c_{4}^{\tau^\prime }-c_8^\tau c_9^{\tau^\prime} \right].
\end{eqnarray}


\providecommand{\newblock}{}

\end{document}